\documentclass[reprint,nofootinbib]{revtex4-2}
\usepackage{graphicx}
\usepackage[
colorlinks=true,
filecolor=black,
anchorcolor=blue,
linkcolor=blue,
citecolor=cyan, 
urlcolor=cyan,
linktocpage=true,
plainpages=false,
breaklinks=true,
pdfstartview=FitH
]{hyperref}

\usepackage{amsmath, amsthm, amssymb,slashed}
\usepackage{color}
\usepackage{multirow}
\usepackage[capitalise]{cleveref}
\usepackage{float}
\usepackage{textgreek}
\usepackage{xspace}

\graphicspath{{Figs},{figs}}

\def\meth{\texttt{\textSigma Track}\xspace}

\begin{document}

\title{Methods for Detecting Gravitational Waves from mini-Extreme-Mass-Ratio Inspirals \uppercase\expandafter{\romannumeral 1}: Statistics Based on Time-Frequency Signal Tracks}
\author{Zi-Xuan Wang$^{1,2}$}
\email{wangzixuan243@mails.ucas.ac.cn}
\author{Gong Cheng$^{1,2}$}
\email{chenggong@ucas.ac.cn}
\author{Ju Chen$^{1,2}$}
\email{chenju@ucas.ac.cn}
\author{Huai-Ke Guo$^{1,2}$}
\email{guohuaike@ucas.ac.cn}
\author{Andrew L. Miller$^{1,2,3,4}$}
\email{andrewlawrence.miller@ligo.org}
\affiliation{$^{1}$International Center for Theoretical Physics Asia-Pacific (ICTP-AP), University of Chinese Academy of Sciences, Beijing 100190, China}
\affiliation{$^{2}$Taiji Laboratory for Gravitational Wave Universe (Beijing/Hangzhou), University of Chinese Academy of Sciences, Beijing 100049, China}
\affiliation{$^{3}$Nikhef - National Institute for Subatomic Physics, Science Park 105, 1098 XG Amsterdam, The Netherlands}
\affiliation{$^{4}$Institute for Gravitational and Subatomic Physics (GRASP), Utrecht University, Princetonplein 1, 3584 CC Utrecht, The Netherlands}

\begin{abstract}

Mini-extreme-mass-ratio inspirals (mini-EMRIs), composed of a stellar-mass compact object and a much lighter companion, are promising sources of continuous gravitational waves in the frequency band of ground-based interferometers such as LIGO-Virgo-KAGRA. Such systems, consisting of sub-solar-mass compact objects, provide a unique probe of exotic compact objects, including primordial black holes. Detecting such long-lived signals, however, remains challenging.
Here, we adapt standard methods used in searches for quasi-monochromatic signals to search for mini-EMRIs, and derive a statistical framework that explicitly handles spectral leakage. In particular, we introduce a new method that sums along the tracks in the time-frequency plane carved out by possible mini-EMRI signals, which we call \texttt{\textSigma Track}.
This refinement establishes a general basis for analyzing long-duration transient signals with rapid frequency evolutions, regardless of the underlying mechanism for gravitational-wave emission.
We also compute a new semi-analytic sensitivity estimate within our new statistical framework, which is valid under the assumption that the signal is weak with respect to the noise level. We then establish a statistic that quantifies how to discretize the search parameter space for our method, which works for mini-EMRIs, as well as arbitrary signal types.
Our results provide a foundation for mini-EMRI searches and demonstrate the potential of current ground-based detectors to probe the existence of sub-solar-mass compact objects.



\end{abstract}

\maketitle











\section{Introduction}


The detection of gravitational waves (GWs) by the 
LIGO-Virgo-KAGRA collaborations \cite{LIGOScientific:2018mvr,LIGOScientific:2020ibl,KAGRA:2021vkt} has driven the rapid development of GW astronomy. Current detections arise from transient burst signals from the mergers of binary black holes, binary neutron stars \cite{LIGOScientific:2016aoc,LIGOScientific:2017vwq,LIGOScientific:2020aai},and neutron star-black hole systems \cite{LIGOScientific:2020zkf,LIGOScientific:2021qlt}. Beyond detecting these short-lived signals, observing long-lasting continuous gravitational waves (CWs), characterized by their persistent and quasi-monochromatic nature, represent one of the next milestones in GW astronomy.

The primary sources of CWs are rotating neutron stars that are asymmetric with respect to their spin axes \cite{LIGOScientific:2020gml,LIGOScientific:2021hvc,LIGOScientific:2021quq,KAGRA:2022dwb,KAGRA:2022osp}. Such asymmetries potentially arise from deformations or ``mountains'' on the neutron star's crust, which result in the emission of persistent, quasi-monochromatic GW signals. 
Beyond neutron stars, similar GW signatures are expected to arise from several exotic processes involving dark-matter candidates. These include the annihilation of boson clouds that could form around spinning black holes via a superradiance process  \cite{Arvanitaki:2014wva,Brito:2015oca,LIGOScientific:2021rnv}, the direct interaction of ultralight dark matter with laser interferometers \cite{Pierce:2018xmy,Guo:2019ker,Miller:2020vsl,Vermeulen:2021epa,LIGOScientific:2021ffg,Miller:2022wxu,Manita:2023mnc,KAGRA:2024ipf,LIGOScientific:2025ttj}, and the inspiral of sub-solar compact objects \cite{LIGOScientific:2018glc,LIGOScientific:2019kan,Miller:2020kmv,LIGOScientific:2021job,Nitz:2021mzz,LIGOScientific:2022hai,Nitz:2022ltl,Miller:2024rca,Miller:2024fpo,LIGOScientific:2025vwc}. Comprehensive overviews of various CW sources, and of probes of dark matter with GWs, can be found in \cite{Piccinni:2022vsd,Haskell:2023yrv} and \cite{Bertone:2019irm,Bertone:2024rxe,Miller:2025yyx}, respectively.

This work focuses on sub-solar compact objects, which are particularly interesting due to their potential primordial origin. Black holes with sub-solar masses are not expected to be produced through conventional stellar-evolution channels. Therefore, these objects are often considered candidates for primordial black holes (PBHs), which are hypothesized to form from the collapse of quantum fluctuations during inflation \cite{Zeldovich:1967lct,Hawking:1971ei,Carr:1974nx}. Detecting such objects would shed light on the evolution of the early Universe. 

PBHs with asteroid- ($10^{-15} \sim 10^{-10} M_\odot$) or planetary- masses ($10^{-7} \sim 10^{-2} M_\odot$) have garnered special attention due to their potential to constitute a significant fraction of dark matter in the Universe \cite{Carr:2020gox,Carr:2020xqk,Bird:2022wvk}. In particular, recent microlensing observations of stars and quasars suggest that planetary-mass PBHs with masses between $10^{-6} M_\odot$ and $10^{-5} M_\odot$ could account for approximately $2\%$ to $10\%$ of dark matter \cite{Niikura:2019kqi,Bhatiani:2019,Hawkins:2020zie}.

Beyond microlensing surveys, GWs offer a promising avenue for detecting sub-solar mass PBHs. These compact objects may form binary systems or pair with stellar-mass black holes or neutron stars, potentially detectable by current ground-based detectors. For asteroid- or planetary-mass PBHs paired with an ordinary compact object, the resulting system would be an extreme-mass-ratio inspiral (EMRI) system \cite{Barsanti:2021ydd,Miller:2024khl}, also referred to as mini-EMRIs \cite{Guo:2022sdd} to distinguish it from traditional EMRIs, which involve ordinary compact objects and supermassive black holes. 

In the sub-solar mass range, the inspiral of binaries would produce steady spin-up signals in the ground-based detector frequency band, lasting from minutes to days to months or even years before merger. While historically searches for sub-solar compact binaries primarily focused on the mass range $[0.1, 1]M_\odot$ \cite{LIGOScientific:2018glc,LIGOScientific:2019kan,LIGOScientific:2021job,LVK:2022ydq,Phukon:2021cus,Nitz:2021mzz,Nitz:2022ltl} using matched filtering, systems with masses below $0.1M_\odot$ remained unexplored until recently \cite{Miller:2024fpo,LIGOScientific:2025vwc}. These signals resemble long-lasting CWs or long-transient signals that fall between canonical CW signals from rotating neutron stars and typical stellar-mass black-hole mergers. Thus, methods originally designed to search for CWs can be adapted to look for sub-solar mass systems. Recently, several techniques, including the (Generalized) Frequency-Hough transform \cite{Miller:2020kmv}, the Viterbi algorithm \cite{Alestas:2024ubs}, and the heterodyne-based BSD-COBI method \cite{Andres-Carcasona:2023zny,Andres-Carcasona:2024jvz}, have been proposed for detecting such signals. Additionally, an approach based on complex-valued STFT maps have also been developed to handle similar long-transient signals \cite{Tenorio:2025gci}. Moreover, GPU-accelerated pipelines have also been proposed to enhance these kinds of searches \cite{Tenorio:2024jgc,Merou:2025ark}. 

These methods have been recently used to constrain PBHs across the planetary-mass regime. In particular, upper limits arising from all-sky searches for CWs in LIGO O3 data have been reinterpreted to constrain the abundance of PBHs \cite{Miller:2021knj,KAGRA:2022dwb}; and, a search using the Generalized Frequency-Hough method has been performed on the same data to look for planetary-mass compact binaries \cite{Miller:2024fpo,Miller:2024jpo,LIGOScientific:2025vwc}. 
For more information on the applicability of CW and matched-filtering methods to sub-solar mass searches, see this recent review \cite{Miller:2024rca}.

These methods exclusively analyze time-frequency spectrograms to look for compact binaries, which have followed the statistical framework described in \cite{Krishnan:2004sv,Palomba:2005fp,Astone:2014esa}. In this work, we revisit this statistical framework, and focus on long-transient signals with moderate frequency evolution, such as mini-EMRIs. We find that spectral leakage alters the statistical properties of the detection statistic when the signal undergoes noticeable frequency evolution within a single Fourier transform (a coherent segment). 

We thus present a revised statistical framework that explicitly accounts for spectral leakage, which provides a general foundation for analyzing long-duration signals with rapid frequency evolutions with respect to those from standard CW sources.


The article is organized as follows. In \cref{sec:signal_model}, we introduce the signal model for mini-EMRIs; in \cref{sec:method_overview}, we provide a brief overview of the commonly-used CW methods, with an emphasis on their statistical frameworks. In \cref{sec:revised_statistics}, we present our new statistics that encapsulate spectral leakage and account for the changing signal amplitude over its duration, along with validations via with simulations of mini-EMRI signals injected into Gaussian noise, and our new method, \meth. In \cref{sec:spectral_leakage}, we provide a detailed treatment of spectral leakage under the weak-signal approximation, and derive new sensitivity estimations in \cref{sec:detection_statistics} for \meth. In \cref{sec:parameter_space_grid}, we discuss how to construct a grid in a search parameter space, and conclude in \cref{sec:conclusion} with directions for future work.



\section{Signal Model}
\label{sec:signal_model}

For inspiraling binary systems, the waveforms for the two GW polarizations in general relativity can be written as
\begin{equation}
\begin{aligned}
    h_+(t) &= h_0(t) \frac{1+\cos^2\iota}{2} \cos\Phi(t),\\
    h_\times(t) &= h_0(t) \cos\iota~ \sin\Phi(t).
    \label{eq:signal_polarization}
\end{aligned}
\end{equation}
where $h_0(t)$ is the amplitude evolution over time $t$, and $\iota$ is 
the inclination angle of the orbit with respect to the line of sight.  $\Phi(t)$ is the phase evolution of the GW: $\Phi(t) \equiv \int^t 2\pi f_{\rm recv}(t') dt'$, where $f_{\rm recv}$ is the frequency received by the GW detector. 
The detector response is then a combination of the two polarizations
\begin{equation}
s(t) = F_+(t) h_+(t) + F_\times(t) h_\times(t),
\end{equation}
where $F_+$ and $F_\times$ are the detector's antenna pattern functions, which vary periodically over one sidereal day. These functions depend on the source's sky location and polarization angle, and the detector's position and orientation on Earth. For simplicity, we omit these angles here; further details can be found in \cite{Jaranowski:1998qm,Riles:2022wwz}. Combining the waveforms given in \cref{eq:signal_polarization}, the detected signal can be reformulated as
\begin{equation}
s(t) = h_0(t) Q(t) \cos\left[\Phi(t) + \phi_p(t)\right],
\label{eq:signal}
\end{equation}
where 
\begin{equation}
Q(t) = \sqrt{{F_+}^2(t) \left(\frac{1+\cos^2\iota}{2}\right)^2 + {F_\times}^2(t) \cos^2\iota},
\end{equation}
encodes the overall amplitude modulation. $\phi_p(t)$ is a phase modulation introduced due to the polarizations
\begin{equation}
\begin{aligned}
\phi_p(t) &= - \arctan \frac{ 2\cos^2\iota~ F_\times(t) }{ (1+\cos^2\iota)~ F_+(t)},\\
\end{aligned}
\end{equation}
In the early inspiral phase, far from merger, the phase evolution of the GW signal can be described by the adiabatic approximation. In particular, for a binary in a quasi-circular orbit with component masses $m_1$ and $m_2$, the spin-up rate of the GW frequency in the Newton approximation is \cite{Maggiore:2007ulw}
\begin{align}
    \dot{f}_{\rm N} = \frac{96}{5} \pi^{8/3} \left(\frac{G\mathcal{M}_c}{c^3}\right)^{5/3} f^{11/3} \equiv k f^{11/3},\label{eq:f_dot}
\end{align}
where $f$ is the GW frequency, and $\mathcal{M}_c \equiv {(m_1 m_2)^{3/5}}{(m_1+m_2)^{-1/5}}$ is the chirp mass of the binary system. The frequency evolution as a function of time follows from integrating \cref{eq:f_dot},
\begin{align}
    f(t) = f_0\left[1 - \frac{8}{3} k f_0^{8/3}(t - t_0) \right]^{-3/8},\label{eq:model_f}
\end{align}
where $f_0$ is the GW frequency at a reference time $t_0$.
The amplitude evolution is
\begin{align}
    h_{0, \rm N}(t) = \frac{4}{d} \left(\frac{G\mathcal{M}_c}{c^2}\right)^{5/3} \left(\frac{\pi f(t)}{c}\right)^{2/3}, \label{eq:model_h}
\end{align}
where $d$ is the distance to the binary.

As the compact objects evolve toward the innermost stable circular orbit (ISCO), relativistic effects become significant, and require numerical relativity to model accurately. For the mini-EMRI systems considered here, such effects can be modeled using perturbation theory. Correction factors for the frequency spin-up rate and amplitude were introduced in \cite{Guo:2022sdd} to \cref{eq:f_dot,eq:model_h}, assuming a circular orbit
\begin{equation}
\begin{aligned}
    \dot{f} &=\dot{f}_{\rm N} ~ C_f(a, f),\\
    h_0(t)&= h_{0, \rm N}(t) ~ C_h(a, f(t)),
    \label{eq:relativistic}
\end{aligned}
\end{equation}
where $\dot{f}$ and $h_0$ is the spin-up rate and amplitude including relativity correction. The factors $C_f$ and $C_h$ can be computed numerically using the Teukolsky formalism.

\begin{figure}
    \centering
    \includegraphics[]{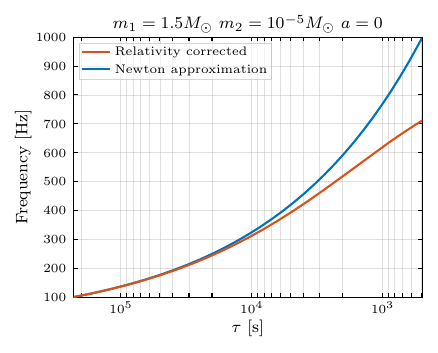}
    \includegraphics[]{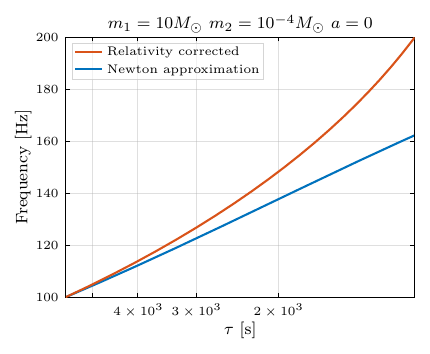}
    \caption{\textbf{Comparison of the frequency evolution of typical mini-EMRI systems with and without relativistic corrections.} The horizontal axis $\tau$ denotes the time to coalescence under the Newtonian approximation. The orbits are assumed to be circular, and the spins of the objects are neglected. The mass ratio is fixed at $q \simeq 10^{-5}$. The upper and lower panels correspond to systems with a neutron star and a stellar-mass black hole as the primary object, respectively. In these two cases, the correction factor is smaller than one in the former and greater than one in the latter.}
    \label{fig:correction}
\end{figure}

A comparison of the frequency evolutions with and without relativistic corrections is shown in \cref{fig:correction}. The upper and lower panels correspond to two mini-EMRI systems with mass ratio $q \sim 10^{-5}$, in which the primary is a typical neutron star and a stellar-mass black hole, respectively. Depending on the system's mass, the correction factor can be either greater or smaller than one.

\begin{figure}
    \centering
    \includegraphics[]{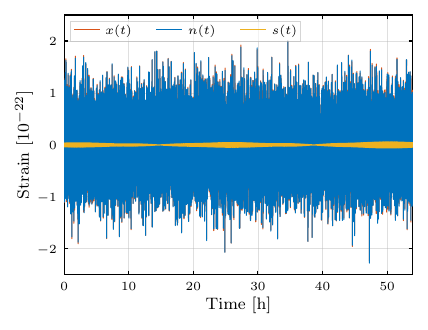}
    \caption{\textbf{Mini-EMRI signal $s(t)$ injected into Gaussian noise $n(t)$. } The system consists of a primary mass of $1.5 M_\odot$ and a companion mass of $10^{-5} M_\odot$ at a distance of $8\,\mathrm{kpc}$. The noise is generated using the O3 LIGO-H1 PSD and band-pass filtered between 100 and 200 Hz. The sky location and orbital orientation are selected randomly. The binary orbit is assumed to be circular, and the spins of the objects are neglected. $x(t)$ is covered completely by $n(t)$.}
    \label{fig:signal}
\end{figure}

\section{Method Overview}
\label{sec:method_overview}

\begin{figure}
    \centering
    \includegraphics[width=\linewidth]{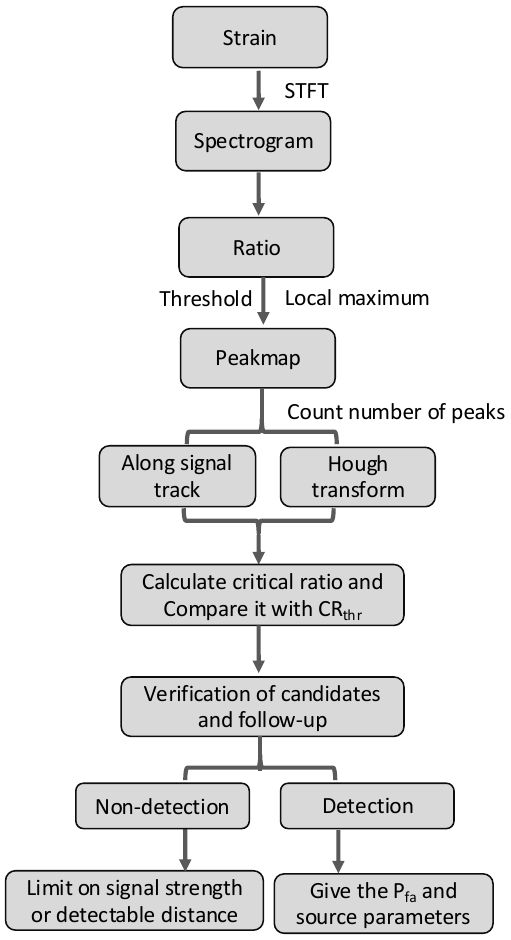}
    \caption{\textbf{The scheme of CW methods based on the Hough transform, but including our new method, \meth, which replaces the Hough transform}.}
    \label{fig:diagram}
\end{figure}


\subsection{Notation and conventions}
Let $x(t)$ be the detector output at time $t$, which is the sum of the detector's response to a gravitational-wave signal $s(t)$ and the detector noise $n(t)$
\begin{equation}
    x(t) = s(t) + n(t).
\end{equation}
Throughout this work, $n(t)$ is assumed to be stationary, Gaussian and have zero mean. As an illustration of the kind of data with which we are working, we show in \cref{fig:signal} the time series for $x(t)$, $s(t)$ and $n(t)$ for a mini-EMRI system, in which the primary object can be considered to be a neutron star with a mass of $1.5M_\odot$.

Throughout this work, we frequently invoke the Discrete Fourier Transform (DFT), and use the convention  
\begin{equation}
\begin{aligned}
&\tilde{x}[k] = \frac{1}{M}\sum_{m=0}^{M-1} x[m] e^{-i\frac{2\pi}{M} mk},\\
&x[m] = \sum_{k=0}^{M-1} \tilde{x}[k] e^{i\frac{2\pi}{M} mk},
\label{eq:DFT}
\end{aligned}
\end{equation}
where $m, k = 0, 1, \ldots, M-1$.  
To simplify subsequent expressions, we adopt the forward normalization by placing the factor $1/M$ in the forward DFT, in contrast to the more common convention.  
In addition, unlike the definition in \cite{Krishnan:2004sv}, the sampling interval $\Delta t$ is not included here.

Normally, CW methods start from the whitened spectrogram, in which the short-time Fourier transform (STFT) of the data, in chunks of duration $T_{\rm DFT}$, is normalized by an estimation of the power spectral density of the noise. The time-series given in \cref{fig:signal} is thus broken into coherent chunks of $T_{\rm DFT}$, and the power is combined across chunks incoherently in CW methods. These methods are thus \emph{semi-coherent}.

Following \cite{Krishnan:2004sv,Astone:2014esa}, we introduce the power ratio $R$ as the basis for constructing the detection statistic
\begin{equation}
    R^{(i)}_k \equiv \frac{|\tilde{x}_i[k]|^2}{\langle|\tilde{n}_i[k]|^2\rangle}, 
    \label{eq:ratio}
\end{equation}
where the subscript $i$ denotes the $i$-th time segment and $k$ denotes the $k$-th frequency bin.  
In practice, the ensemble average in the denominator is often estimated using an autoregressive average over previous time segments to account for potential non-stationarity in the noise; see \cite{Astone:2005fj} for details.  
In this work, for simplicity, we assume the noise to be stationary.

As shown in \cite{Krishnan:2004sv,Astone:2014esa}, twice the power ratio, $2R$, follows a non-central $\chi^2$ distribution with two degrees of freedom
\begin{equation}
    2R \sim \chi^2\left(2, \lambda\right),
    \label{eq:R_chi2_distribution}
\end{equation}
where the non-centrality parameter $\lambda$, which now keeps track of time segment and frequency bin labels, is defined as
\begin{equation}
    \lambda^{(i)}_k \equiv 2\,\frac{|\tilde{s}_i[k]|^2}{\langle|\tilde{n}_i[k]|^2\rangle}.
    \label{eq:lambda_ik_def}
\end{equation}
$\lambda$ represents twice the ratio of signal power to noise power, i.e., twice the signal-to-noise ratio (SNR) in the corresponding time-frequency pixel.
In the limit that the discrete Fourier transform can be approximated by its continuous version, assuming negligible spectral leakage such that the signal power is confined within a single frequency bin, the statistic for the signal-dominated bin can be written as
\begin{equation}
    \lambda_\mathrm{ideal} = \frac{T_\mathrm{DFT}h_0^2Q^2}{S_n(f)},
    \label{eq:lambda_approx}
\end{equation}
where $S_n(f)$ denotes the single-sided power spectral density (PSD) of the detector noise. A detailed derivation is provided in \cref{sec:lambda}.

It is important to distinguish this theoretical quantity, $\lambda_\mathrm{ideal}$, from the actual statistic values, denoted as $\lambda_k^{(i)}$, encountered in practical data analysis. Here, $\lambda_\mathrm{ideal}$ serves only as a reference value under idealized approximations; in contrast, the observed $\lambda_k^{(i)}$ values are distributed across the time-frequency map and are subject to spectral leakage.

The probability density function of $R$ is then given by
\begin{equation}
     p(x; \lambda) = e^{-x - \frac{\lambda}{2}} I_0\!\left(\sqrt{2 \lambda x}\right),
     \label{eq:R_distribution}
\end{equation}
where $x$ is an arbitrary value of $R$, and $I_0$ denotes the modified Bessel function of the first kind and zeroth order.  
In the absence of a signal, $\lambda$ vanishes, and $2R$ follows a central $\chi^2(2)$ distribution.  
The corresponding distribution of $R$ reduces to
\begin{equation}
     p(x; 0) = e^{-x}.
     \label{eq:R_distribution_noise}
\end{equation}





\subsection{Statistics framework}

\subsubsection{Overview}
Here, we first outline the general data-analysis pipeline of a standard CW semi-coherent method, the Hough transform \cite{Krishnan:2004sv, Astone:2014esa}. The scheme of the method is shown in \cref{fig:diagram}.
In this pipeline, the time-series strain data are first band-pass filtered and downsampled. 
The data are then divided into short segments and transformed into a time-frequency map, or \emph{spectrogram}, via the discrete STFT. 
The spectrogram is subsequently whitened according to \cref{eq:ratio}, producing a ratio map from which pixels exceeding the threshold $R \ge \theta$ are selected.
In a particular implementation of the Hough transform, the FrequencyHough \cite{Astone:2014esa}, an additional local-maximum criterion is applied, resulting in a \emph{peakmap}.


The remaining pixels are accumulated along linear signal tracks given by parameter values in the search space (the slope and y-intercept of these lines, i.e. $\dot{f}$ and $f_0$, respectively). Tracks with counts exceeding a predefined threshold $n_{\rm th}$ are identified as signal candidates.
This procedure can be efficiently implemented using the Hough transform \cite{Krishnan:2004sv, Astone:2014esa, Miller:2018rbg,Menon:2025wce,Miller:2025ote}, through direct demodulation with optimized parameter-space searches \cite{Andres-Carcasona:2023zny}, or with machine learning \cite{Miller:2019jtp,Alestas:2024ubs}.
Each selected candidate is then cross-checked against known instrumental noise lines and validated through coincidence analysis of two detectors' data. If candidates pass these tests, the follow-up is performed, in which the data are demodulated for particular signal parameters. This demodulation makes the signal more monochromatic and permits the use of longer DFT segments, thus improving the sensitivity, or allowing the candidate to be rejected if it is not astrophysical. 



\subsubsection{Hough-transform statistics}

We now turn to describe how the input to Hough-transform searches, the peakmap, is constructed. First, the spectrogram is built in steps of duration $T_{\rm DFT}$, which is chosen to ensure that the signal's frequency evolution remains confined within half a frequency bin,  
$\delta f = \dot{f} T_{\rm DFT} < \Delta f/2$, where $\Delta f = 1/T_{\rm DFT}$ is the the frequency resolution of the DFT \cite{Krishnan:2004sv, Astone:2014esa}:
\begin{equation}
    T_{\rm DFT} < \sqrt{\frac{1}{2|\dot{f}|_{\max}}},
    \label{eq:T_DFT}
\end{equation}
where $|\dot{f}|_{\max}$ is the maximum frequency derivative of the particular signal searched for.

Following \cite{Astone:2014esa}, for a time-frequency peak to appear in the peakmap, the corresponding ratio $R$ must exceed both the threshold value $\theta$ and the values of $R$ in thet two adjacent bins.
Given the distribution of $R$ described in \cref{eq:R_chi2_distribution,eq:R_distribution_noise}, the probability that a peak appears in the peakmap due to a signal is \cite{Astone:2014esa, Palomba:2025}
\footnote{Note that there is an error in the definition of $p_\lambda$ given in \cite{Astone:2014esa},  
\begin{equation}
    p_\lambda = \int_\theta^{+\infty} p(x; \lambda) 
    \left[\int_0^x p(x'; \lambda)\, dx'\right]^2 dx,
\end{equation}
as well as in the small-signal approximation presented in Eq.~(20) of the same reference.  
The presence of $\lambda$ in the inner integral is a typo. A correction is provided in 
\cite{Palomba:2025}, following our presentation in an internal discussion.}
\begin{equation}
    p_\lambda = \int_\theta^{+\infty} p(x; \lambda)
    \left[\int_0^x p(x'; 0)\, dx'\right]^2 dx.
    \label{eq:p_lambda}
\end{equation}
Here, the signal is assumed to be fully confined within a single frequency bin, with negligible contribution to adjacent bins.

In the absence of a signal, the probability that a peak appears in the peakmap is
\begin{equation}
     p_0 = \int_\theta^{+\infty} p(x; 0) \left[\int_0^x p(x';0) dx'\right]^2 dx.
     \label{eq:p_0}
\end{equation}



Furthermore, the noise processes in different segments can be considered independent, which means that the accumulated peak counts in the peakmap, $n$, along any track defined by parameters in the parameter space, follow a \emph{binomial distribution},
\begin{equation}
  n \sim B(N, p),
\end{equation}
where $p = p_\lambda$ for tracks that align with that of a a real signal, and $p = p_0$ for tracks that do not correspond to any signal.
For large $N$, the binomial distribution can be effectively approximated by a normal distribution, with mean and variance given by
\begin{equation}
  \mu_n = Np, \quad \sigma_n^2 = Np(1-p).
  \label{eq:normal_approx}
\end{equation}
This approximation is valid provided that $N$ is large and $p$ is not too close to 0 or 1.

Signal candidates can be selected by applying a threshold on the peak counts along a track. However, the threshold depends on the noise properties. To isolate the contribution of noise and make the statistic independent of noise variations, Refs. \cite{Krishnan:2004sv,Astone:2014esa} introduced the \emph{critical ratio} (CR)
\begin{equation}
    \mathrm{CR} \equiv \frac{n - Np_0}{\sqrt{Np_0 \left(1 - p_0\right)}}.
\end{equation}
Here, the random peak counts $n$ are normalized by the mean and variance of the noise-only case. In the absence of a signal, the CR follows a standard normal distribution with zero mean and unity standard degviation. In the presence of a signal, however, the CR deviates from the standard normal distribution, providing a measure of the statistical significance of $n$ for any given trajectory in the peakmap.

Signal candidates are then selected by applying a threshold on the CR, determined by a predefined false-alarm probability—that is, the probability that a candidate arises from noise alone. If no detection is confirmed after follow-up verification of candidates, a detection limit can be set based on a predefined probability that a true signal fails to exceed the threshold, i.e., the false-dismissal probability. Further details are provided in \cref{sec:detection_statistics}.

\section{\meth: Revised Statistics}
\label{sec:revised_statistics}
\begin{figure}
    \centering
    \includegraphics[]{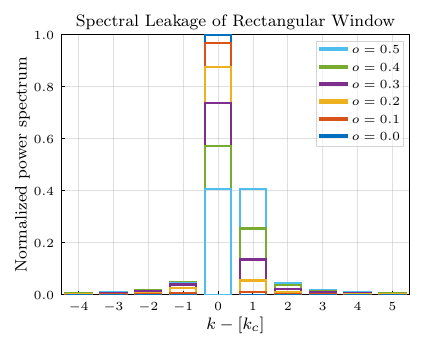}
    \caption{\textbf{The relative power contained in different frequency bins for a monochromatic signal which has been windowed by a rectangular window a rectangular window}. Different colors correspond to different offsets $o$ of the signal frequency from the frequency bin in which it should fall. The horizontal axis is bin indices, referenced to the signal frequency. This figure is adapted from \cite{Allen:2002bp,Riles:2022wwz}.}
    \label{fig:leakage}
\end{figure}

In the previous statistical framework, spectral leakage was not considered; however, it is inevitable due to the finite DFT length.  
To illustrate this effect, we first consider a monochromatic signal and take a DFT of duration $T_{\rm DFT}$, corresponding to a frequency resolution of $ f_\mathrm{bin} = 1/T_{\rm DFT}$. If the frequency of the signal matches a muliple of a frequency bin, then one bin will accumulate all of the signal power; otherwise,
any mismatch introduces spectral power loss across nearby bins, which can be as large as 50\%, as shown in \cref{fig:leakage} and detailed further in \cref{sec:leakage}.

In previous studies \cite{Krishnan:2004sv,Astone:2014esa}, a correction factor was introduced for $\lambda$ in \cref{eq:p_lambda} to account for the power loss in the central bin.
For a monochromatic signal with a rectangular window, the DFT component $\tilde{s}_k$ corresponds to discrete samples of the sinc function, which is the Fourier transform of the rectangular window. If we consider the signal frequency relative to the frequency bin it occupies, we can write $f - f_k \in [-\tfrac{1}{2}, \tfrac{1}{2}] f_\mathrm{bin}$. With that, and noting that $\lambda$ is proportional to $|\tilde{s}_i[k]|^2$, the averaged correction factor is
\begin{equation}
    \int_{-\frac{1}{2}}^{\frac{1}{2}} {\rm sinc}^2(o)\, {\rm d} o = 0.7737,
\end{equation}
where ${\rm sinc}\,(o) = {\sin(\pi o)}/{\pi o}$ and $o = (f - f_k)/{f_\mathrm{bin}}$ denotes the fractional bin offset of the signal.

However, the signal contribution to adjacent bins was still ignored, which is not appropriate when spectral leakage is considered. Moreover, unlike the monochromatic case, even under the requirement of \cref{eq:T_DFT}, the signal's frequency evolution can still extend across two frequency bins. This can be easily understood by shifting the initial frequency of a signal within a given DFT time segment. In fact, a non-negligible fraction of time segments fall into this category.

It can be verified that the probability distribution of $2R$, for any given time-frequency pixel in the spectrogram, remains a $\chi^2$ distribution, even in the presence of spectral leakage. However, the signal is no longer fully confined to a single frequency bin, as assumed previously, due to both spectral leakage and the intrinsic signal's frequency evolution. Consequently, the probability that a peak appears in the peakmap, given in \cref{eq:p_lambda,eq:p_0}, should be modified as
\begin{equation}
\begin{aligned}
    p_{\lambda}(i,k) &=& \int_\theta^{+\infty} p(x; \lambda_k^{(i)}) 
    \left[\int_0^x p(x'; \lambda_{k-1}^{(i)}) \, dx'\right] \\
    && \times \left[\int_0^x p(x'; \lambda_{k+1}^{(i)}) \, dx'\right] dx,
    \label{eq:p_lambda_k}
\end{aligned}
\end{equation}
where $\lambda_k^{(i)}$ and $\lambda_{k\pm1}^{(i)}$ represent the noise-normalized signal power in the central and adjacent frequency bins in time segment $i$.

One point to note is that even if the signal amplitude is assumed constant, the signal frequency evolution still leads to a variable frequency span across different time segments, which alters the spectral leakage and thus the normalized power $\lambda_k^{(i)}$. Consequently, $p_\lambda$ varies along the signal track, i.e. it changes in different time segments (see \cref{sec:spectral_leakage}). In this case, the peak counts along tracks defined by parameters in the parameter space no longer follow a simple binomial distribution. Each pixel along the track is a realization of a Bernoulli distribution with a slightly different $p_\lambda$. The total count is therefore the sum of $N$ independent Bernoulli random variables, with mean and variance given by
\begin{equation}
\begin{aligned}
    \mu_n & = \sum_{i=1}^{N} p_{\lambda}(i,k), \\
    \sigma^2_n & = \sum_{i=1}^{N} p_{\lambda}(i,k) \left[1 - p_{\lambda}(i,k)\right].
    \label{eq:normal_approx_modified}
\end{aligned}
\end{equation}
When $N$ is sufficiently large, according to the law of large numbers, the distribution of the total peak count $n$ can be well approximated by a normal distribution $n\sim \mathcal{N}(\mu_n, \sigma_n)$.


With the modified distribution of counts $n$ along the signal track, the CR can be defined same as before, but with a new mean and variance
\begin{equation}
\begin{aligned}
    \mu_{\mathrm{CR}} & =  \frac{\mu_n-Np_0}{\sqrt{N p_0 \left(1 - p_0\right)}}, \\
    \sigma^2_{\mathrm{CR}} & = \frac{\sigma_n^2}{Np_0(1-p_0)}.
    \label{eq:normal_approx_modified_2}
\end{aligned}
\end{equation}
To compute these new statistics in practice, we introduce \meth, which sums the peaks along the time-frequency tracks of mini-EMRI signals.

We verified the accuracy of \cref{eq:normal_approx_modified_2} with \meth using simulated signals injected into Gaussian noise. The signal parameters are identical to those in \cref{fig:signal}, except that the amplitude is fixed for easier comparison with the previous formula. The signal frequency evolves from $f_{\min} = 100~\mathrm{Hz}$ to $f_{\max} = 200~\mathrm{Hz}$ over a total duration of 194280~s. The data are divided into 8~s DFT segments, tapered with a flat-cosine window and 50\% overlap between adjacent segments. The signal amplitude is normalized such that $\mathcal{L} = 1$, where $\mathcal{L} \equiv 2P/\langle |\tilde{n}[k]|^2\rangle$ formally defined later in \cref{eq:total_power_lambda}.

\begin{figure}[h]
    \centering
    \includegraphics[]{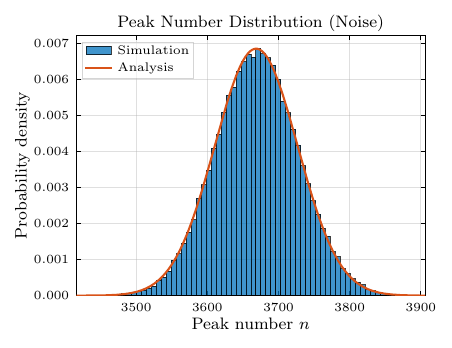}
    \includegraphics[]{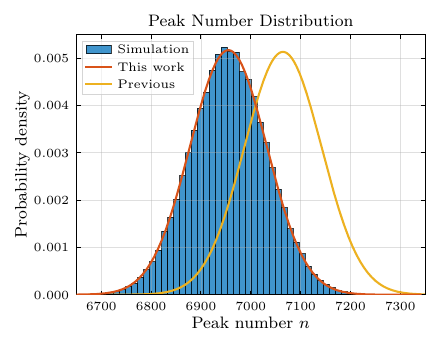}
    \caption{
    \textbf{Probability density distribution of peak counts $n$ for noise (upper panel) and signal (lower panel) tracks.} The blue histograms show the simulation results, while the orange and red curves represent the distributions modeled by \cref{eq:normal_approx,eq:normal_approx_modified}, respectively.
    }
    \label{fig:counts_distribution}
\end{figure}

Using the same signal with different noise realizations, we simulated the distribution of peak counts $n$ along the tracks in the peakmap.  
\cref{fig:counts_distribution} shows the resulting distributions for both the signal and non-signal tracks.  
For the signal track (lower panel), the simulated distribution closely follows the normal distribution with mean and variance given by \cref{eq:normal_approx_modified}, where the values of $\lambda_k^{(i)}$ in \cref{eq:p_lambda_k} are obtained directly from the simulation.  
In contrast, the predictions from \cref{eq:normal_approx} show a systematic bias, even after applying a correction factor to account for power loss in the central-bin.  
In the pure-noise case (non-signal track, upper panel), the two formulas yield identical results and are plotted as a single red curve.

\begin{figure}
    \centering
    \includegraphics[]{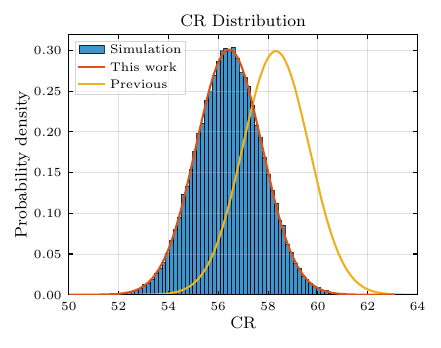}
    \includegraphics[]{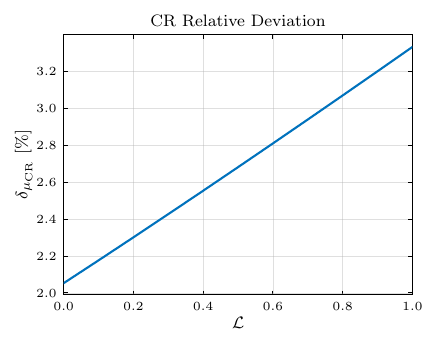}
    \caption{\textbf{Probability density function for the critical ratio CR (upper panel) and the relative deviation (in percent) of the mean CR between the previous and new statistical approaches (lower panel)}. In the upper panel, the blue histogram represents the simulation results, and the orange and red curves correspond to the theoretical distributions modeled by \cref{eq:normal_approx} and \cref{eq:normal_approx_modified}, respectively. The lower panel displays the relative deviation $\delta_{\mu_\mathrm{CR}}=(\mu_\mathrm{CR}^\prime - \mu_\mathrm{CR}) /\mu_\mathrm{CR}$ as a function of $\mathcal{L}$, representing the bias of the neighbor-neglecting approximation ($\mu_\mathrm{CR}^\prime$) relative to the full theoretical prediction ($\mu_\mathrm{CR}$) derived in this work.
    }
    \label{fig:CR_distribution}
\end{figure}

The distributions of the corresponding CRs for the signal track, obtained from two simulations using both statistical frameworks, are shown in the upper panel of \cref{fig:CR_distribution}.  
Again, the updated statistical model exhibits much better agreement with the simulation.  
We further examined the relative deviation of the mean CR between the two frameworks for various signal strengths and found that the deviation increases with signal strength, as shown in the lower panel of \cref{fig:CR_distribution}.

Note that both the orange and red curves in \cref{fig:counts_distribution} and \cref{fig:CR_distribution} are obtained from \meth, i.e. summing along the exact signal track, not the generalized frequency-Hough transform, as used in the previous work. Compared to the generalized frequency-Hough transform, \meth can be more sensitive because we count exactly one pixel in a given time segment, corresponding to where the signal is located at that time. However, this is not always guaranteed in the generalized frequency-Hough transform. A fixed pixel in the Hough map corresponds to a belt in the peakmap, and all the peaks in this belt will vote for that pixel in the Hough map. Because of the signal's frequency evolution, the generalized frequency-Hough transform would count two pixels or none for some segments. We have tested this for the simple power-law signal, evolving from $100 ~{\rm Hz}$ to $200 ~{\rm Hz}$ with duration of about $2.5 \times 10^5$ seconds, and have seen that this miscounting leads to a loss of approximately $61\%$ segments, which corresponds to a loss in the $\text{CR}$ of $\sim38\%$. Note that the loss fraction decreases for narrower frequency bands. This phenomenon arises from projecting a uniform frequency grid onto a non-uniform grid of $x = 1/f^{n-1}$, and therefore affects only the frequency-Hough transform, not the traditional Hough transform.

\section{Spectral Leakage Analysis}
\label{sec:spectral_leakage}
Under the weak signal approximation, where $\lambda_k$ and $\lambda_{k\pm1}$ are small, \cref{eq:p_lambda_k} can be linearized as
\begin{equation}
    p_\lambda(i,k) = p_0\left[1 + m\lambda^{(i)}_{k} + n\left(\lambda^{(i)}_{k+1} + \lambda^{(i)}_{k-1}\right)\right],
\end{equation}
where $m$ and $n$ are functions of the threshold $\theta$ (detailed expressions are provided in \cref{sec:weak_signal_approximation}). To encapsulate the first-order effects of $\lambda^{(i)}_k$ and its neighboring bins, we define a new statistic $\Lambda$, such that for each grid point in the time-frequency map,
\begin{equation}
    \Lambda^{(i)}_k = m\lambda^{(i)}_k + n\left(\lambda^{(i)}_{k+1} + \lambda^{(i)}_{k-1}\right).
    \label{eq:Lambda}
\end{equation}
The mean value of $\Lambda$ is then
\begin{equation}
    \bar{\Lambda} = \frac{1}{N}\sum_{i=1}^{N}\Lambda^{(i)}_k.
\end{equation}
Accordingly, the expected value and variance of the random variable CR are
\begin{equation}
\begin{aligned}
    \mu_{\mathrm{CR}} &= \sqrt{\frac{N p_0}{1-p_0}}\,\bar{\Lambda},\\
    \sigma_{\mathrm{CR}}^2 &= 1 + \frac{1-2p_0}{1-p_0}\,\bar{\Lambda}.
    \label{eq:muCR_sigmaCR}
\end{aligned}
\end{equation}

The distribution of the statistic $\lambda^{(i)}_k$ across the time-frequency map can be factorized into a total-power term and a normalized power-distribution term,
\begin{equation}
\lambda^{(i)}_k = \frac{2P_i}{\langle|\tilde{n}_i[k]|^2\rangle} \cdot \frac{|\tilde{s}_i[k]|^2}{P_i}.
\label{eq:total_power_and_eta}
\end{equation}
The first factor, defined as the total power statistic $\mathcal{L}_i$, quantifies the average signal power within the $i$th time segment relative to the local noise PSD:
\begin{equation}
\mathcal{L}_i \equiv \frac{2 P_i}{\langle|\tilde{n}_i[k]|^2\rangle}.
\label{eq:total_power_lambda}
\end{equation}
Here, $P_i$ denotes the mean signal power over the analysis window:
\begin{equation}
    P_i \equiv \frac{1}{M}\sum_{m=0}^{M-1}|s_i[m]|^2 = \sum_{k=0}^{M-1}|\tilde{s}_i[k]|^2,
\end{equation}
where the second equality follow from Parseval's theorem.

The second factor, $\eta^{(i)}_k$, is the normalized power spectrum, which characterizes how the signal's power is distributed across frequency bins $k$ within the $i$th time segment:
\begin{equation}
\eta_k^{(i)} = \frac{|\tilde{s}_i[k]|^2}{P_i}.
\label{eq:eta_k}
\end{equation}
When a window function is applied, $\tilde{s}_i[k]$ should be appropriately normalized so that the total power remains unchanged, which ensures that the normalized power spectrum $\eta^{(i)}_k$ satisfies the unity condition $\sum_{k=0}^{M-1} \eta^{(i)}_k = 1$.



\begin{figure}
    \centering
    \includegraphics[]{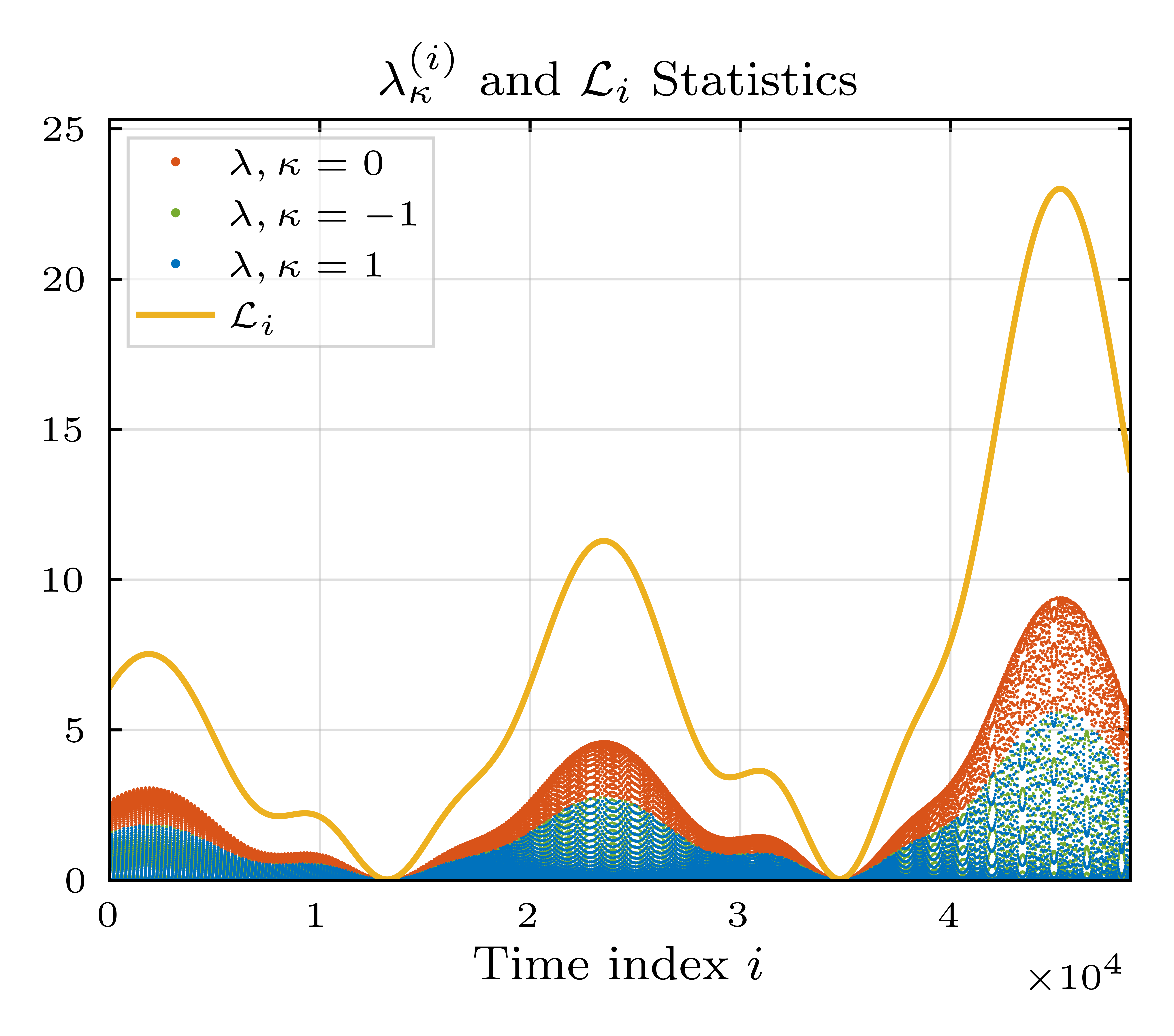}
    \caption{\textbf{Signal power plotted over time in the frequency bin that contains the signal ($\kappa=0$) and two adjacent bins ($\kappa=\pm 1)$}. The red dots in the figure mark the values of $\lambda^{(i)}_{\kappa=0}$ at the true signal frequency, while the blue and green dots denote the adjacent bins $\lambda^{(i)}_{\kappa=\pm 1}$, respectively. The yellow curve shows the total-power statistic $\mathcal{L}_i$ defined in \cref{eq:total_power_lambda}. For visual clarity, we smooth the O3 LIGO–H1 noise PSD before computing $\mathcal{L}_i$, which removes spurious features and short-timescale fluctuations that would otherwise obscure the underlying trend.
    Here the signal parameters are the same as those used in \cref{fig:signal}. The DFT segment duration is set to $T_{\mathrm{DFT}} = 8\,\mathrm{s}$, using a Tukey window with $\alpha = 0.5$ and a $50\%$ overlap between adjacent segments.
    }
    \label{fig:lambda_and_lambda_ik_with_response}
\end{figure}

For a general real-valued, frequency-modulated signal, the power distribution in the frequency domain is characterized by two distinct features: the symmetric bipartition of energy between positive and negative frequencies, and the local spectral leakage around the central frequency $f_c$ driven by the signal's evolution. As discussed in \cref{sec:lambda}, in the simplest theoretical model where local leakage is negligible, $\lambda_\mathrm{ideal}$ captures only the positive frequency component. Consequently, it represents half of the total signal capacity, leading to the fundamental relation $\lambda_\mathrm{ideal} = \mathcal{L}/2$. And, the total power statistic $\mathcal{L}_i$ can be approximated as:
\begin{equation}
    \mathcal{L}_i \approx \frac{2T_{\mathrm{DFT}}\,h_{0,i}^2\,Q_i^2}{S_{n,i}}.
    \label{eq:L_approx}
\end{equation}
The magnitude of $\mathcal{L}_i$ varies with the time index $i$ due to three primary factors: the secular evolution of the amplitude $h_{0,i}$, the diurnal modulation of the detector response $Q_i$, and the variation in the local noise PSD $S_{n,i}\equiv S_n(f_c^{(i)})$ as the signal frequency evolves. To validate this analysis, we performed a simulation with the same signal parameters used in \cref{fig:signal}. The results, illustrating the relationship between the recovered $\lambda$ and the theoretical $\mathcal{L}$, are shown in \cref{fig:lambda_and_lambda_ik_with_response}.


\begin{figure}
    \centering
    \includegraphics[]{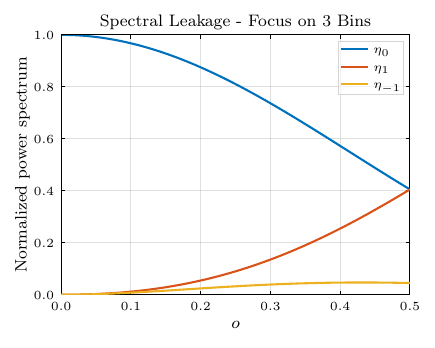}
    \caption{\textbf{Signal power in the bin coincident with the central frequency of the signal, as well as in the adjacent bins, plotted as a function of the fractional bin offset.} As the signal frequency increasily does not align with a frequency bin, the signal power becomes more and more spread to neighboring bins.
        }
    \label{fig:leakage_continue}
\end{figure}

The normalized power spectrum, $\eta_k$, as defined in \cref{eq:eta_k}, describes the power distribution across the entire frequency domain. When focusing on the frequency band containing the signal, $\eta_k$ can be interpreted as a measure of spectral leakage. To formalize this perspective, we introduce a new indexing convention centered on the signal's frequency. For a given segment in the time-frequency analysis, let the signal's central frequency be $f_c$. 
We convert this to a frequency-bin index, $k_c = f_c / f_{\text{bin}}$, which is not necessarily an integer. 
The primary bin containing the signal is then identified by the index $\lfloor k_{c} \rceil$, where $\lfloor x \rceil$ denotes a special rounding function that rounds to the nearest integer but leaves half-integers unchanged (i.e., $\lfloor x \rceil = x$ if $x$ is a half-integer). 
A detailed definition of this function is provided in \cref{sec:leakage}.

\begin{figure}
    \centering
    \includegraphics[]{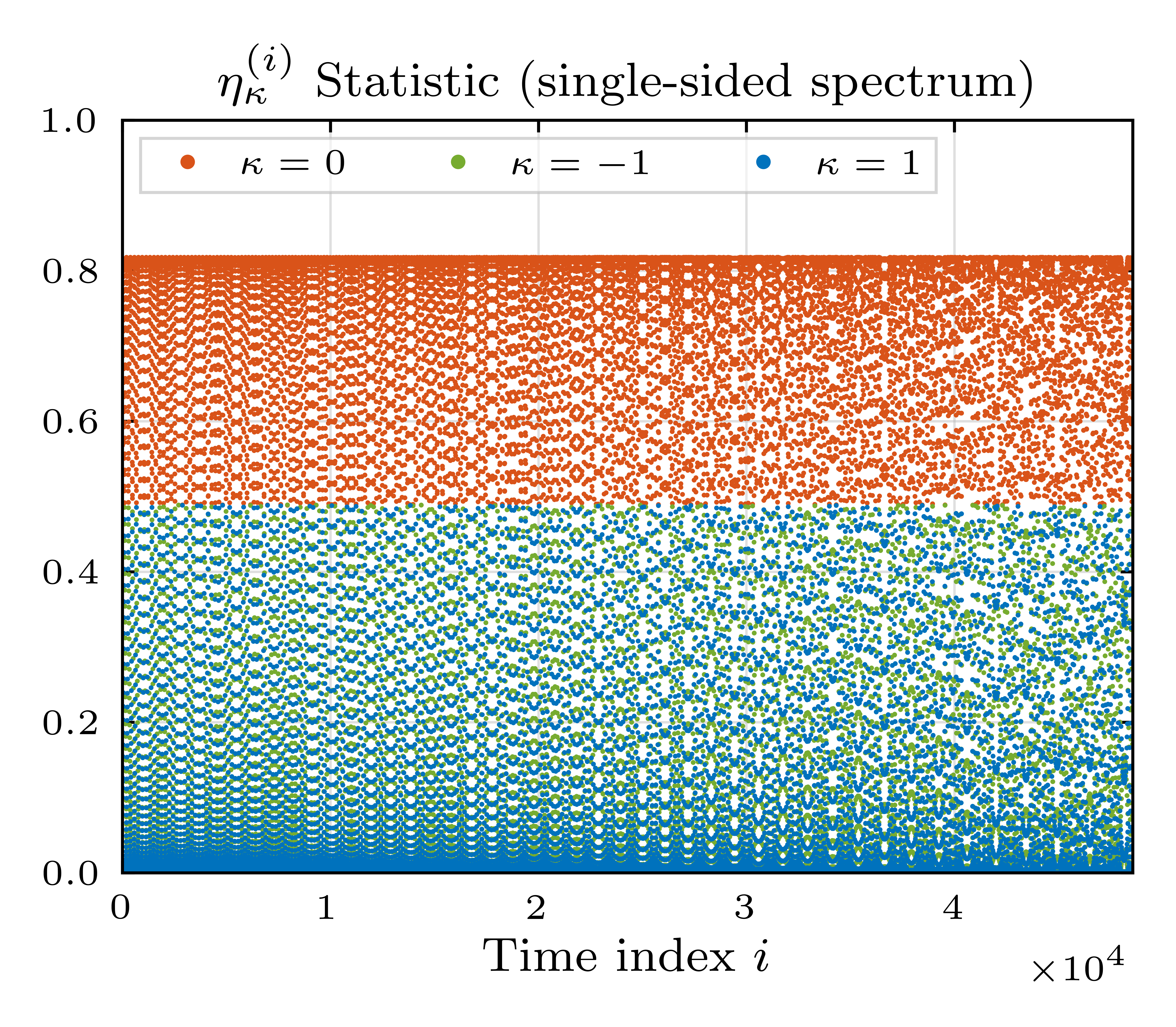}
    \caption{\textbf{The spectral leakage of the signal into adjacent frequency bins.} The red, blue, and green dots represent the normalized power in the primary signal bin ($\eta^{(i)}_0$) and the adjacent lower and upper frequency bins ($\eta^{(i)}_{-1}$ and $\eta^{(i)}_{+1}$), respectively. The values are doubled to represent a single-sided power spectrum for real-valued signal. This figure is obtained by dividing $\lambda^{(i)}_{\kappa=0,\pm 1}$ by $\mathcal{L}_i$ in \cref{fig:lambda_and_lambda_ik_with_response}.}
    \label{fig:three_eta}
\end{figure}

To quantify the power distribution around the signal, we introduce a signal-anchored index, $\kappa$:
\begin{equation}
    \kappa = k - \lfloor k_{c} \rceil.
\end{equation}
In this convention, $\eta_0$ represents the fraction of power within the signal's primary bin, while $\eta_{\pm 1}$ correspond to the power leaked into the adjacent frequency bins. For the signal example shown in \cref{fig:lambda_and_lambda_ik_with_response}, we compute the spectral leakage components using \cref{eq:total_power_and_eta}. These components are plotted as a function of time in \cref{fig:three_eta}, revealing that their values oscillate rapidly.

\begin{figure}
    \centering
    \includegraphics[]{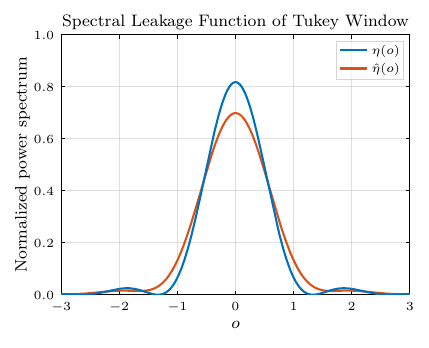}
    \caption{\textbf{The normalized power spectrum corresponding to the spectral leakage function $\eta(o)$ and its average $\hat{\eta}(o)$ employed in this work}. Both functions are plotted as a function of the offset between the signal frequency and the frequency bin in which it falls. These functions are obtained by using the Tukey window ($\alpha = 0.5$)}
\label{fig:Spectral_leakage_comparison}
\end{figure}

To understand this oscillatory behavior, we model the leakage with a continuous spectral leakage function, $\eta(o)$, which is plotted in \cref{fig:Spectral_leakage_comparison}.  
The argument of this function is the frequency offset factor, $o$, defined for each bin $\kappa$ as
\begin{equation}
    o_\kappa = \frac{f_\kappa - f_c}{f_{\text{bin}}},
\end{equation}
which measures the distance from the central frequency $f_c$ in units of bin widths. The discrete leakage factors, $\eta_\kappa$, can thus be understood as integer samples of this continuous function. The rapid oscillations seen in \cref{fig:three_eta} arise because the signal's central frequency, $f_c$, evolves over time. Consequently, the frequency offset factor at each time segment $i$, denoted $o_\kappa^{(i)}$, varies, causing it to sample different points along the continuous spectral leakage function, $\eta(o)$.

\begin{figure}
    \centering
    \includegraphics[]{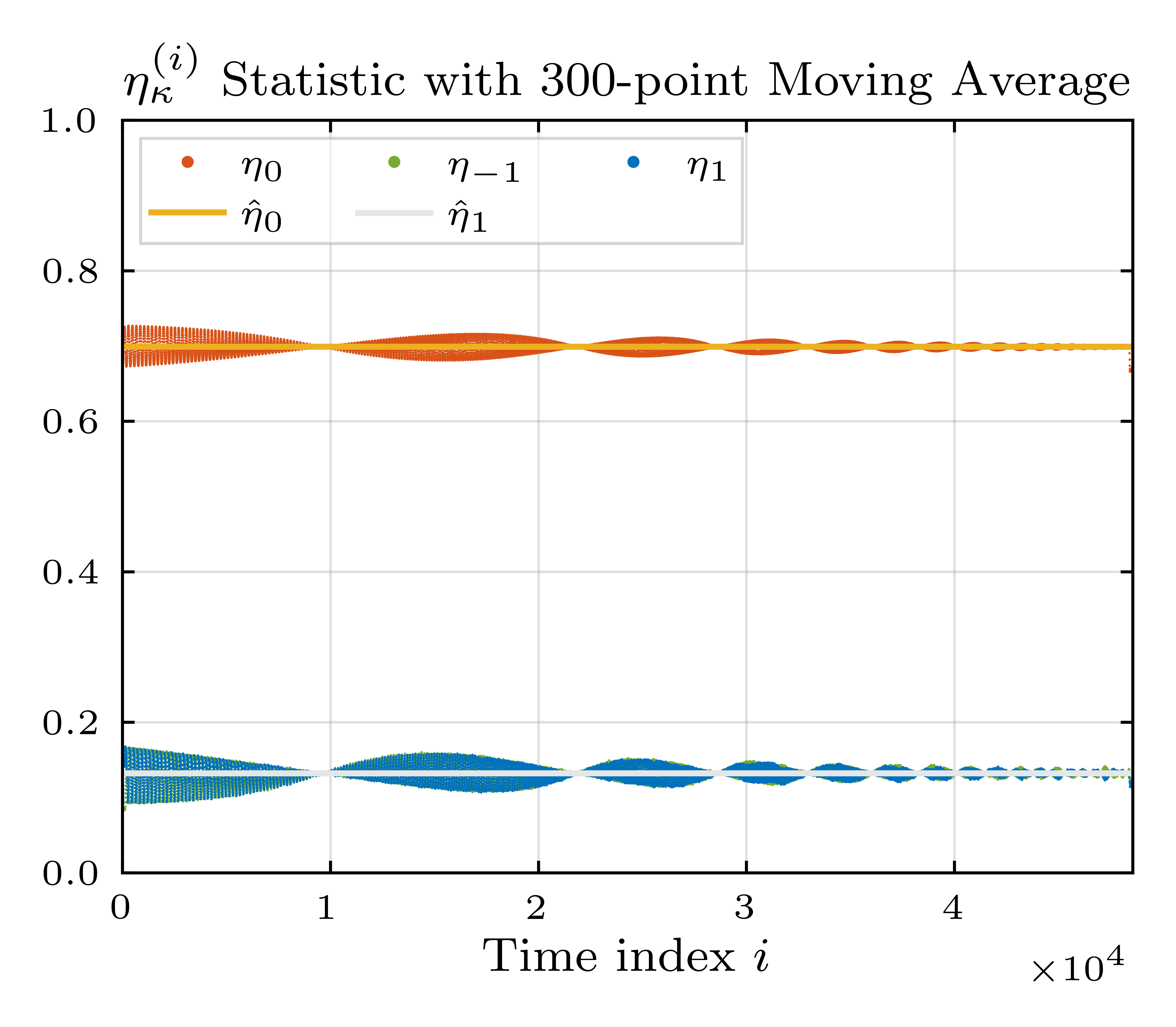}
    \caption{\textbf{Estimation of the average of $\eta_\kappa^{(i)}$ within one frequency bin from the signal frequency, over time}. These averages are calculated using $\eta_{\kappa}^{(i)}$ obtained in \cref{fig:three_eta}. To compute these curves, we calculate the moving average of $\eta_{\kappa}^{(i)}$ over 300 time segments and compare it with the average spectral leakage factor $\hat{\eta}_\kappa$. We find excellent agreement.}
    \label{fig:three_eta_move_average}
\end{figure}

Under the weak-signal approximation, the statistic depends linearly on $\lambda$, allowing the averaging of $p_\lambda$ to be applied directly to $\lambda$. To characterize the average behavior of spectral leakage and obtain a measure less sensitive to rapid fluctuations, we introduce the average spectral leakage function, $\hat{\eta}(o)$:
\begin{equation}
    \hat{\eta}(o) = \int_{o-1/2}^{o+1/2} \eta(o^\prime) \, \mathrm{d}o^\prime.
\end{equation}
This expression represents a moving average of the continuous spectral leakage function, $\eta(o)$, over a window of unit width. From this, we define the average spectral leakage factors, $\hat{\eta}_\kappa$, by sampling the continuous function at integer offsets: $\hat{\eta}_\kappa = \hat{\eta}(\kappa)$.  
Since $\hat{\eta}(o)$ inherits the even symmetry of $\eta(o)$, only non-negative integer values of $\kappa$ need to be considered. The first three theoretical values for several common window functions are listed in \cref{tab:combined_eta_H}. To verify this model, we applied a moving average to the empirical data from \cref{fig:three_eta}. A comparison between the averaged data and the theoretical values from \cref{tab:combined_eta_H} is shown in \cref{fig:three_eta_move_average}, demonstrating excellent agreement.

\begin{table*}[t]
    \centering
    \caption{Comparison of the averaged spectral leakage factor $\hat{\eta}_\kappa$ and the combined factor $\hat{\mathrm{H}}_\kappa$ for different window functions. The values for $\hat{\mathrm{H}}_\kappa$ are calculated with a threshold $\theta = 2.5$. Only the first three orders are shown.}
    \label{tab:combined_eta_H}
    \setlength{\tabcolsep}{12pt} 
    \renewcommand{\arraystretch}{1.2} 
    \begin{tabular}{l c c c c c c}
    \hline\hline
    & $\hat{\eta}_0$ & $\hat{\eta}_1$ & $\hat{\eta}_2$ & $\hat{\mathrm{H}}_0$ & $\hat{\mathrm{H}}_1$ & $\hat{\mathrm{H}}_2$ \\ 
    \hline
    Rectangular             & 0.7737 & 0.07870 & 0.01403              & 0.9738 & 0.05008 & 0.01248 \\
    Tukey ($\alpha{=}0.5$)  & 0.6991 & 0.1322  & 0.01577              & 0.8721 & 0.1227  & 0.01154 \\
    Hanning                 & 0.6009 & 0.1969  & 0.002599             & 0.7390 & 0.2120  & -0.009191 \\
    Hamming                 & 0.6466 & 0.1755  & 0.001039             & 0.7998 & 0.1821  & -0.009815 \\
    Bartlett                & 0.6578 & 0.1691  & 0.0006619 & 0.8149 & 0.1732  & -0.009958 \\
    Blackman &0.5339 & 0.2215 & 0.01157 &  0.6508 & 0.2470  & 0.0006637 \\
    \hline\hline
    \end{tabular}
\end{table*}

The statistical distributions of interest (e.g., for quantities $n$ and CR) are governed by the time-averaged quantity, $\bar{\Lambda}$, which represents the mean of $\Lambda_k^{(i)}$ over all time segments. As derived in \cref{sec:leakage}, when $\mathcal{L}_i$ varies slowly, this average can be expressed as
\begin{equation}
    \bar{\Lambda} = \frac{1}{N}\sum_{i=1}^{N} \mathcal{L}_i \, \hat{\mathrm{H}}(o),
    \label{eq:bar_Lambda_eta}
\end{equation}
where $\hat{\mathrm{H}}(o)$ is the average combined spectral leakage function. It is defined as a weighted sum of the average spectral leakage function, $\hat{\eta}(o)$:
\begin{equation}
    \hat{\mathrm{H}}(o) = m \, \hat{\eta}(o) + n \left[ \hat{\eta}(o+1) + \hat{\eta}(o-1) \right],
\end{equation}
with coefficients $m$ and $n$ derived from the series expansion of the detection statistic under the weak-signal approximation. The construction of this combined quantity, $\hat{\mathrm{H}}(o)$, follows the same motivation as the definition of $\Lambda_k^{(i)}$.

\section{Detection Statistics}
\label{sec:detection_statistics}

Based on the statistical distribution of the CR, a threshold can be set corresponding to a specified false-alarm probability. Signal candidates exceeding this threshold are then subject to follow-up analysis. In the absence of a detection, a detection limit is established by additionally specifying a false-dismissal probability. In this section, we derive the theoretical upper limit on the maximum detectable distance. 


We can express the false-alarm probability $P_{\mathrm{fa}}$ and false-dismissal probability $P_{\mathrm{fd}}$ in terms of the CR and its mean and standard deviations as
\begin{equation}
\begin{aligned}
P_{\mathrm{fa}} &= \int_{\mathrm{CR_{thr}}}^{\infty} p_{\mathrm{CR,\,noise}}\, d{\rm CR} = \frac{1}{2}\,{\rm erfc}\!\left(\frac{{\rm CR_{thr}}}{\sqrt{2}}\right),\\[3pt]
P_{\mathrm{fd}} &= \int_{-\infty}^{\mathrm{CR_{thr}}} p_{\mathrm{CR,\,signal}}\, d{\rm CR} = \frac{1}{2}\,{\rm erfc}\!\left(\frac{\mu_{\mathrm{CR,\,signal}} - {\rm CR_{thr}}}{\sqrt{2}\,\sigma_{\mathrm{CR,\,signal}}}\right).
\end{aligned}
\end{equation}
These relations can be rearranged to yield a constraint on the expected value of the CR for detectable signals:
\begin{equation}
    \mu_{\mathrm{CR}} \geq \sqrt{2}\,\sigma_{\mathrm{CR}}\,\mathrm{erfc}^{-1}\!\left(2P_{\mathrm{fd}}\right)
    + \sqrt{2}\,\mathrm{erfc}^{-1}\!\left(2P_{\mathrm{fa}}\right).
\end{equation}
Using the weak-signal approximation given in \cref{eq:muCR_sigmaCR}, this expression can be further simplified to obtain the minimum required value of the time-averaged quantity $\bar{\Lambda}$,

\begin{equation}
    \bar{\Lambda} \geq \sqrt{\frac{1-p_0}{Np_0}} \left[ \sqrt{2}~\mathrm{erfc}^{-1}(2P_{\mathrm{fd}}) + \sqrt{2}~\mathrm{erfc}^{-1}(2P_{\mathrm{fa}}) \right].
    \label{eq:Lambda_min}
\end{equation}
Here, the $\bar{\Lambda}$ term in $\sigma_{\mathrm{CR}}$ is neglected, since for a sufficiently large number of segments $N$, 
$\sqrt{\tfrac{N p_0}{1-p_0}} \gg 
\sqrt{2}\,\mathrm{erfc}^{-1}(2P_{\mathrm{fd}})\tfrac{1-2p_0}{1-p_0}$.

Following \cref{eq:bar_Lambda_eta}, for a time-frequency track that perfectly follows the signal evolution, we have
\begin{equation}
    \bar{\Lambda} \approx \hat{\mathrm{H}}_0\,\frac{1}{N}\sum_{i=1}^{N} \mathcal{L}_i,
    \label{eq:Lambda_bar_sum}
\end{equation}
where $\mathcal{L}_i = 2 h_{0,i}^2 Q_i^2 T_{\mathrm{DFT}}/S_{n,i}$, and $\hat{\mathrm{H}}_0$ denotes the value of $\hat{\mathrm{H}}(o)$ with zero offset $o=0$. Further details can be found in \cref{sec:averaged_slf}.
To derive the constraint on the detectable distance, we factor the distance out of the signal amplitude $h_0$ by introducing a distance-independent amplitude $h_{1\mathrm{kpc}} = h_0 \, (d / {\rm kpc})$.
Combining \cref{eq:Lambda_min,eq:Lambda_bar_sum}, the maximum detectable distance is obtained as

\begin{align}
d_{\mathrm{max}} &= \left( \frac{N p_0}{1 - p_0} \right)^{1/4}  \left( \mathcal{A} ~ T_{\mathrm{DFT}} ~ \hat{\mathrm{H}}_0 \right)^{1/2} \label{eq:d_max} \\
&\times \left[ \sqrt{2}\,\mathrm{erfc}^{-1}(2P_{\mathrm{fa}}) + \sqrt{2}\,\mathrm{erfc}^{-1}(2P_{\mathrm{fd}}) \right]^{-1/2} ~{\rm kpc}, \nonumber
\end{align}
\noindent where $\mathcal{A} = \frac{1}{N} \sum_{i=1}^{N} h_{1\mathrm{kpc},i}^2 Q_i^2/S_{n,i}$.

The quantity $\hat{\mathrm{H}}_0 \sim m\,\hat{\eta}_0^{1/2}$ when leakage into neighboring bins is ignored, where $m = C_m/p_0$ as defined in \cref{eq:m_n_def}, and $C_m$ corresponds to $p_1$ in previous work~\cite{Palomba:2025}. Neglecting variations in the signal amplitude due to intrinsic evolution and detector response, as well as changes in the local noise power spectral density arising from frequency evolution, this expression reduces to

\begin{align}
d_{\mathrm{max}} =& h_{1\mathrm{kpc}} ~\langle Q\rangle~\hat{\eta}_0^{1/2} \frac{N^{1/4} T_{\mathrm{DFT}}^{1/2} }{ S_n^{1/2} }
\left[ \frac{p_1^2}{p_0(1 - p_0)} \right]^{1/4} \label{eq:d_max_simplified}\\
& \times \left[ \sqrt{2}\,\mathrm{erfc}^{-1}(2P_{\mathrm{fa}}) + \sqrt{2}\,\mathrm{erfc}^{-1}(2P_{\mathrm{fd}}) \right]^{-1/2} ~{\rm kpc}. \nonumber
\end{align}

Noting that $\sqrt{2}\mathrm{erfc}^{-1}(2P_{\mathrm{fd}}) = {\rm CR_{thr}}$ and $\mathrm{erfc}^{-1}(2P_{\mathrm{fa}}) = -\mathrm{erfc}^{-1}(2\Gamma)$, where $\Gamma = 1 - P_{\mathrm{fa}}$ is the confidence level defined in~\cite{Palomba:2025}, this expression is fully consistent with the $h_{0,\min}$ result of Eq.~(29) in~\cite{Palomba:2025}, upon taking 
$\langle Q \rangle = 2/5$ and $\hat{\eta}_0^{1/2} = \sqrt{2.4308/\pi}$ for a rectangular window.

\section{Parameter-space grid}
\label{sec:parameter_space_grid}
\begin{figure*}[hbt] 
    \centering
    \includegraphics[]{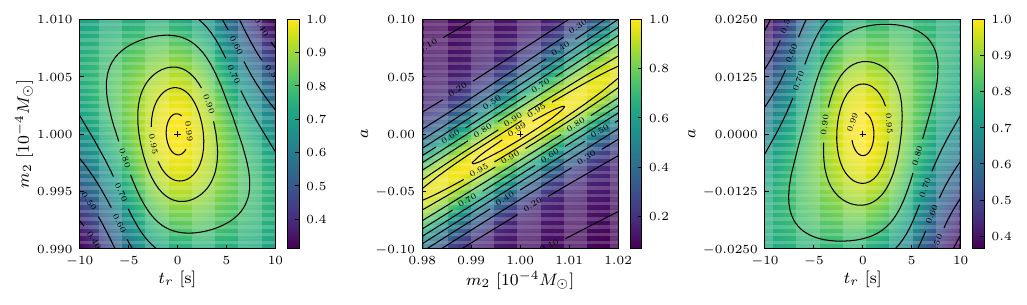}
    \caption{
   \textbf{Coordinate-plane slices of the FF distribution in the three-dimensional parameter space $(m_2, a, t_r)$ for a mini-EMRI system.} 
   The reference frequency is set to $f_r = 170\,\mathrm{Hz}$ to minimize correlations between $t_r$ and the physical parameters.
   }
    \label{fig:Lambda_bar_LR_miniEMRI}
\end{figure*}

\meth allows us to follow more complicated time-frequency trajectories, such as those that arise from mini-EMRIs, in contrast to the Hough transform.
Although this increases the computational cost, GPU-accelerated computation is expected to mitigate the burden, and the sensitivity should improve through a more accurate match to the signal model. Here, we briefly introduce the construction of the parameter-space grids, which is the first step in performing a search with \meth for mini-EMRI signals.

The intrinsic waveform of a mini-EMRI signal is mainly determined by four parameters: $m_1, m_2, a, C$ \cite{Guo:2019ker}. Here, we ignore the eccentricity $e$ and consider only circular orbits. To fully describe the mini-EMRI signal, seven additional parameter are required, including the sky location $\alpha, \delta$, the orientation angles $\iota, \psi$, the distance $d$, and the time $t_r$ corresponding to a predefined reference frequency $f_r$.

However, for \meth, we only need to consider the signal track in the time-frequency map. The distance or amplitude modulation due to variations of the detector's antenna pattern does not affect the shape of the trajectory, and thus can be omitted when discretizing the search parameter space. Moreover, the frequency splitting of a monochromatic signal due to the diurnal motion of the earth, encapsulated in the polarization phase modulation $\phi_p(t)$, can also be ignored, because this splitting is small compared to the intrinsic frequency change of the mini-EMRI signal. That means that the only remaining parameters are ($\alpha, \delta$), the sky position of the source, which causes a Doppler shift due to the orbital motion of the earth around the sun with respect to the source location.
However, for a targeted or directed search, the sky location is known or constrained within a small region, thus the Doppler effect can be well modeled. In contrast, for an all-sky search, the sky position would need to be included in the search grid to account for Doppler modulations. For min-EMRI signals considered in this work that have rapid intrinsic frequency evolution, the Doppler effect becomes smaller than the frequency shift induced by the inspiral itself; thus, the sky grid can be coarser, or even one point, depending on $T_{\rm DFT}$ and the frequency \cite{Astone:2014esa}.

On the other hand, the intrinsic frequency evolution is mainly determined by the masses of the system ($m_1, m_2$) and the spin of the primary ($a$). Here, the compactness parameter $C$ is omited in our initial consideration. When the signal is far from the merger, the relativistic correction $C_f$ in \cref{eq:relativistic} can be ignored. The mass parameters $m_1, m_2$ are degenerate, and the wave waveform is only determined by the combined chirp mass $\mathcal{M}_c$, and is also insensitive to $a$. As the signal evolves close to the ISCO, the relativistic correction would break the mass degeneracy, and the waveform becomes sensitive to $a$.

In a directed search, in which we target a specific neutron star or black hole, we except the mass of the primary to be well constrained. The search grid would then focus on three parameters: $m_2, a$ and the time $t_r$ that corresponds to given frequency $f_r$.

To rigorously design the search grid, we must quantify how the detection statistic degrades due to the mismatch between a template and a real signal. We define the track-averaged statistic $\bar{\Lambda}(\mathcal{P}_t|\mathcal{P}_r)$, where $\mathcal{P}_r$ and $\mathcal{P}_t$ denote the true signal parameters and the template parameters, respectively,
\begin{equation}
\begin{aligned}
    \bar{\Lambda}(\mathcal{P}_t|\mathcal{P}_r) &= \frac{1}{N}\sum_{i=1}^{N} \mathcal{L}_i \, \hat{\mathrm{H}}\left(o^{(i)}\right).
\end{aligned}
\end{equation}
The offset factor $o^{(i)}$ in the $i$-th time segment is the frequency difference between the template's centre frequency $f_{c,t}^{(i)}$, and that of the real signal $f_{c,r}^{(i)}$, normalized by the frequency bin size
\begin{equation}
    o^{(i)} = \frac{f_{c,t}^{(i)} - f_{c,r}^{(i)}}{f_{\mathrm{bin}}}.
\end{equation}
The search efficiency is characterized by the Fitting Factor (FF), defined as the ratio of the recovered statistic to the optimal statistic obtained by a perfectly matched template:
\begin{equation}
    \mathrm{FF}(\mathcal{P}_t|\mathcal{P}_r) \equiv \frac{\mu_\mathrm{CR}(\mathcal{P}_t|\mathcal{P}_r)}{\mu_\mathrm{CR}(\mathcal{P}_r|\mathcal{P}_r)} \approx \frac{\bar{\Lambda}(\mathcal{P}_t|\mathcal{P}_r)}{\bar{\Lambda}(\mathcal{P}_r|\mathcal{P}_r)}.
    \label{eq:fitting_factor}
\end{equation}
The approximation on the right holds under the weak-signal limit.

To align with our sensitivity metric (maximum detectable distance), we define the Mismatch (MM) as:
\begin{equation}
    \mathrm{MM} \equiv 1 - \sqrt{\mathrm{FF}},
    \label{eq:mismatch_def}
\end{equation}
so that the mismatch directly represents the fractional loss in detection horizon. The goal of template-bank construction is then to tile the parameter space such that the maximum mismatch for any plausible signal  does not exceed a prescribed tolerance (e.g., $\mathrm{MM} \le 3\%$).

In \cref{fig:Lambda_bar_LR_miniEMRI}, we illustrate the morphology of the FF within the three-parameter space spanned by $\{m_2, a, t_r\}$ space using three coordinate-plane slices, with component masses $m_1 = 10\,M_\odot$ and $m_2 = 10^{-4}\,M_\odot$, and spin $a = 0$ as the fiducial template parameters. With the template parameters at the centre of each slice, the contours depict the FF values obtained for systems with varying parameters (frequency range $[100, 200]\,\mathrm{Hz}$). In the high-match regime, the FF contours form regular ellipses, consistent with the local quadratic (Taylor) expansion of the FF around the template parameters.




The reference frequency is set to $f_r = 170\,\mathrm{Hz}$. It is worth noting that this choice is deliberate: it effectively decouples the reference time $t_r$ from the intrinsic physical parameters ($m_2$ and $a$). Geometrically, this decoupling aligns the principal axes of the metric ellipses with the coordinate axes in both the $m_2$--$t_r$ and $a$--$t_r$ planes, thereby minimizing correlations between timing and physical parameters.
This choice is reflected in the left and right panels, where the FF contours align with the coordinate axes, indicating that the time parameter is effectively decoupled from $m_2$ and $a$.


\section{Conclusion}
\label{sec:conclusion}

In this work, we develop \meth, a semi-coherent search for mini-EMRIs, which are a new class of continuous gravitational-wave signals for ground-based detectors. These systems, consisting of a stellar-mass compact object and a much lighter companion, provide a unique probe of exotic compact objects, including primordial black holes.

Following the formalism of standard Hough-transform methods \cite{Krishnan:2004sv,Astone:2014esa,Miller:2020kmv}, we refine their statistical frameworks by explicitly accounting for spectral leakage. Simulation results show deviations from previous treatments, while remaining well described by our updated formulation. We also provide updated semi-analytic sensitivity estimates under the weak-signal approximation and offer a brief discussion of parameter-grid construction. Our results establish a foundation for mini-EMRI searches and demonstrate the potential of current ground-based detectors to probe sub-solar-mass compact objects.

One point to note is that, although the statistical framework is general, we focus primarily on Gaussian noise in this work. Future efforts will incorporate realistic detector noise spectra and time-varying antenna patterns. This will enable more accurate sensitivity estimates and inform search strategies for mini-EMRIs in actual gravitational-wave data.

\begin{acknowledgments}

This material is based upon work supported by NSF's LIGO Laboratory which is a major facility fully funded by the National Science Foundation.

We acknowledge the Python scientific ecosystem, particularly \texttt{NumPy} \cite{harris2020array}, \texttt{SciPy} \cite{2020SciPy-NMeth}, and \texttt{Matplotlib} \cite{Hunter:2007}, which were used in the analysis and visualization.
This work is supported by the National Natural Science Foundation of China (NSFC) under Grant No. 12347103 and 12547104.
\end{acknowledgments}

\appendix

\section{Discussion about $\lambda$}
\label{sec:lambda}

Under the DFT convention adopted in this work, Parseval's theorem is expressed as
\begin{equation}
    P_i = \frac{1}{M}\sum_{m=0}^{M-1} |s_i[m]|^2 = \sum_{k=0}^{M-1} |\tilde{s}_i[k]|^2.
\end{equation}
Consider a quasi-monochromatic signal $s(t)$ whose power is effectively concentrated within a single frequency bin $k^\prime$. The discrete spectral power can be written as
\begin{equation}
    \left|\tilde{s}_i[k]\right|^2 \approx \left(\frac{\delta_{k,k^\prime}}{2} + \frac{\delta_{k,M-k^\prime}}{2}\right)P_i,
\end{equation}
where the total instantaneous power is given by
\begin{equation}
    P(t) = \frac{1}{2}h_0^2(t)Q^2(t).
\end{equation}
Consequently, the power contribution in the positive frequency bin $k^\prime$ is
\begin{equation}
    \left|\tilde{s}_i[k^\prime]\right|^2 \approx \frac{1}{4}h_0^2(t)Q^2(t).
    \label{eq:power_signal}
\end{equation}

Conversely, for a stationary random noise process $n(t)$, the single-sided PSD, $S_n(f)$, is related to the autocorrelation function via the Wiener-Khinchin theorem:
\begin{equation}
    \frac{1}{2}S_n(f) = \int_{-\infty}^{\infty} \langle n(t)n(t+\tau) \rangle e^{-i2\pi f \tau} \,d\tau,
\end{equation}
where $\langle \cdot \rangle$ denotes the ensemble average. Note that for a stationary process, the autocorrelation $\langle n(t)n(t+\tau) \rangle$ depends solely on the time lag $\tau$. In the frequency domain, this implies the orthogonality of Fourier components:
\begin{equation}
    \langle \tilde{n}(f) \tilde{n}^*(f') \rangle = \delta(f-f') \frac{1}{2}S_n(f),
\end{equation}
where $\tilde{n}(f)$ denotes the continuous Fourier transform of $n(t)$. Transiting to the discrete domain, the expected noise power in a frequency bin is approximated by
\begin{equation}
   \langle|\tilde{n}[k]|^2\rangle \approx \frac{S_n(f_k)}{2T_\mathrm{DFT}}.
   \label{eq:power_noise}
\end{equation}

Combining the signal power from \cref{eq:power_signal} and the noise background from \cref{eq:power_noise}, the expectation value of the statistic $\lambda$ in the signal-dominated bin is approximately
\begin{equation}
    \lambda \approx \frac{T_\mathrm{DFT}h_0^2Q^2}{S_n(f)}.
    \label{eq:lambda_approx_app}
\end{equation}
This expression is consistent with the result derived in \cite{Palomba:2005fp}, up to a factor of two arising from a different normalization definition of $\lambda$, while it differs from the conventions adopted in \cite{Krishnan:2004sv,Astone:2014esa}.

Using the stationary phase approximation, the squared magnitude of the continuous Fourier transform of the signal relates to the instantaneous power $P(t)$ via $|\tilde{s}(f)|^2 \approx \frac{1}{2}\dot{f}^{-1}P(t(f))$.
Substituting this into the definition of the statistic, for a signal with slowly varying amplitude, we obtain:
\begin{equation}
    \lambda \approx \frac{4\,|\tilde{s}(f)|^2}{S_n(f)}\,T_\mathrm{DFT}\dot{f}
    = \frac{4\,|\tilde{s}(f)|^2}{T_\mathrm{DFT} S_n(f)}\,\left( \frac{\Delta f}{f_\mathrm{bin}} \right),
    \label{eq:lambda_SPA}
\end{equation}
where $\tilde{s}(f) = \int_{-\infty}^{\infty} s(t)e^{-i2\pi f t}\,dt$ denotes the continuous Fourier transform,
$\Delta f = \dot{f}T_\mathrm{DFT}$ represents the frequency drift of the signal within a single DFT time segment,
and $f_\mathrm{bin} = 1/T_\mathrm{DFT}$ is the DFT frequency resolution.

For comparison, the expression adopted in \cite{Krishnan:2004sv,Astone:2014esa} is
\begin{equation}
    \lambda = \frac{4\,|\tilde{s}(f)|^2}{T_\mathrm{DFT} S_n(f)}.
    \label{eq:lambda_previous}
\end{equation}
The two forms are generally inconsistent and coincide only when $\Delta f = f_\mathrm{bin}$, which requires $T_\mathrm{DFT} = \dot{f}^{-1/2}$.  
More importantly, as the coherent time $T_\mathrm{DFT}$ increases, $\lambda$, which represents the SNR within the signal frequency bin, should increase accordingly.  
However, the expression in \cref{eq:lambda_previous} predicts the opposite behaviour, decreasing with larger $T_\mathrm{DFT}$.

\section{Weak-signal approximation}
\label{sec:weak_signal_approximation}

In this Appendix we derive compact expressions for the peak-selection probability and for the statistics of the peak count along a signal track under the weak-signal approximation. The weak-signal expansion is obtained by Taylor-expanding the exact expression for $p_\lambda(i,k)$ (\cref{eq:p_lambda_k}) to first order in the non-centrality parameters $\lambda^{(i)}_{k}$ and $\lambda^{(i)}_{k\pm1}$.

\subsection{Local maxima and threshold}
\label{sec:local_maxima_and_threshold}
When the non-centrality parameters are small, the peak-selection probability for the grid point $(i,k)$ can be linearized as
\begin{equation}
    p_{\lambda}(i,k) = p_0 + C_{m}\,\lambda^{(i)}_{k}
    + C_{n}\,\big(\lambda^{(i)}_{k+1}+\lambda^{(i)}_{k-1}\big) + \mathcal{O}(\lambda^2),
    \label{eq:weak_linear_p_lambda}
\end{equation}
where $p_0$ denotes the noise-only peak probability and the coefficients $C_m$ and $C_n$ depend only on the threshold $\theta$. Explicitly,
\begin{equation}
\begin{aligned}
    &p_0 = e^{-\theta}-e^{-2 \theta}+\frac{1}{3} e^{-3 \theta},\\
    &C_{m}=\frac{1}{4}e^{-2 \theta}-\frac{ 1}{9}e^{-3 \theta}+\frac{1}{6} e^{-3 \theta} \theta-\frac{1}{2} e^{-2 \theta} \theta+\frac{1}{2}e^{-\theta} \theta, \\
    &C_{n}=\frac{1}{18}e^{-3 \theta}-\frac{1}{8}e^{-2 \theta}+\frac{1}{6} e^{-3 \theta} \theta-\frac{1}{4} e^{-2 \theta} \theta.
\end{aligned}
\label{eq:Cm_Cn}
\end{equation}

It is convenient to factor out $p_0$ and introduce the normalized coefficients
\begin{equation}
    m \equiv \frac{C_m}{p_0},\quad
    n \equiv \frac{C_n}{p_0},
    \label{eq:m_n_def}
\end{equation}
so that \cref{eq:weak_linear_p_lambda} can be written as
\begin{equation}
    p_\lambda(i,k) = p_0\Big[1 + m\,\lambda^{(i)}_{k} + n\big(\lambda^{(i)}_{k+1}+\lambda^{(i)}_{k-1}\big)\Big].
    \label{eq:weak_p_lambda_mn}
\end{equation}

Two limiting cases recover earlier results: if one sets $\lambda^{(i)}_{k}=\lambda^{(i)}_{k\pm1}=\lambda$, the expression reduces to the form used in \cite{Astone:2014esa}; if the neighbor contributions are neglected $\lambda^{(i)}_{k\pm1}=0$ the result coincides with \cite{Palomba:2025}.

\begin{figure}[htb]
    \centering
    \includegraphics[]{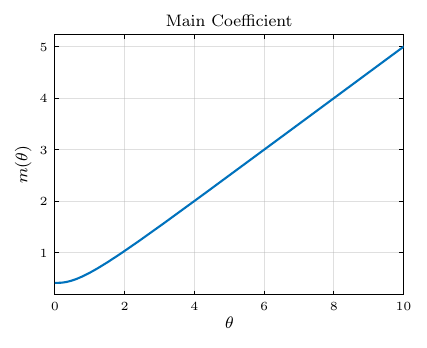}
    \includegraphics[]{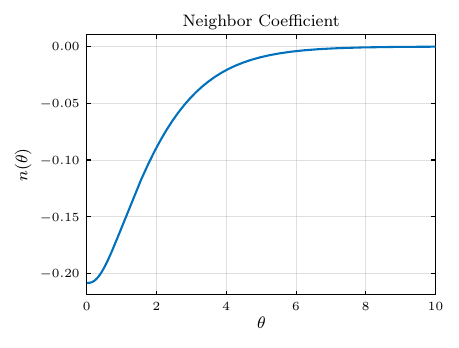}
    \caption{The normalized linear coefficients $m(\theta)$ (upper) and $n(\theta)$ (lower) appearing in \cref{eq:weak_p_lambda_mn}.}
    \label{fig:m_n_functions}
\end{figure}

The functions $m(\theta)$ and $n(\theta)$ are shown in \cref{fig:m_n_functions}. For large $\theta$, the coefficient $m$ depends approximately linearly on $\theta$, whereas $n$ has a much smaller magnitude and tends to zero as $\theta$ increases. Typically $n<0$, indicating that a power contribution in neighbor bins reduces the probability of the central bin forming a local maximum. For example, at $\theta=2.5$ one finds $m\approx1.272$ and $n\approx-0.06345$, so $|n/m|\approx5\%$.

Using the first-order statistic
\begin{equation}
    \Lambda^{(i)}_k \equiv m\,\lambda^{(i)}_{k} + n\big(\lambda^{(i)}_{k+1}+\lambda^{(i)}_{k-1}\big),
    \label{eq:Lambda_first_order}
\end{equation}
and summing along a signal track of $N$ segments with \meth, the mean and variance of the peak count $n$ on that track are
\begin{equation}
\begin{aligned}
    \mu_n  &= N p_0\big(1+\bar{\Lambda}\big),\\[4pt]
    \sigma^2_n &= N\bar{p}(1-\bar{p}) - N p_0^2 \sigma^2_{\Lambda},
    \label{eq:mu_n_sigma_n_app}
\end{aligned}
\end{equation}
where
\begin{equation}
    \bar{\Lambda} \equiv \frac{1}{N}\sum_{i=1}^N \Lambda^{(i)}_k,\quad
    \sigma^2_\Lambda \equiv \frac{1}{N}\sum_{i=1}^N\big(\Lambda^{(i)}_k-\bar{\Lambda}\big)^2.
\end{equation}
Since $\sigma^2_{\Lambda}$ is of second order in the small quantities $\lambda^{(i)}_k$, it may be neglected in a conservative first-order treatment.

Under the same approximation, the CR statistic admits the linear expansion
\begin{equation}
\begin{aligned}
    \mu_{\mathrm{CR}} &= \sqrt{\frac{N p_0}{1-p_0}}\,\bar{\Lambda},\\[4pt]
    \sigma^2_{\mathrm{CR}} &= 1 + \frac{1-2p_0}{1-p_0}\,\bar{\Lambda},
    \label{eq:muCR_sigmaCR_app}
\end{aligned}
\end{equation}
which provides the shift and variance modification of the CR distribution induced by a weak signal.

\begin{figure}[htb]
    \centering
    \includegraphics[]{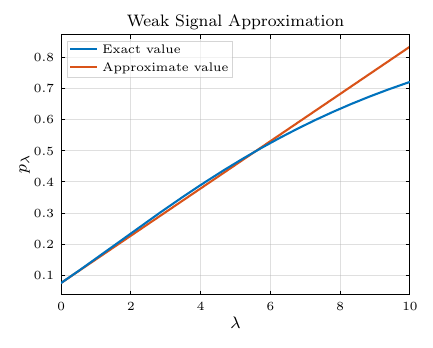}
    \includegraphics[]{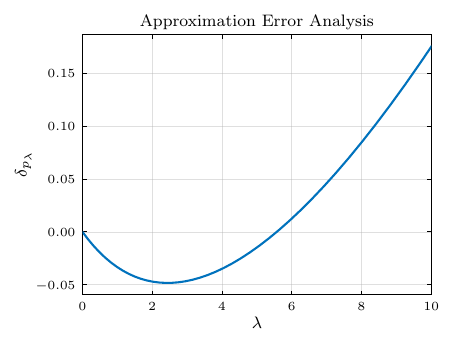}
    \caption{Upper panel: comparison between the exact $p_\lambda$ and the weak-signal linear approximation (example: $\eta_0=0.8$, $\eta_{\pm1}=0.1$, $\theta=2.5$). Lower panel: relative deviation of the approximation, defined as $\delta_{p_\lambda}=(p'_\lambda-p_\lambda)/(p_\lambda-p_0)$, which is proportional to the deviation of CR.}
    \label{fig:weak_signal_approximation}
\end{figure}

The accuracy of the linear (weak-signal) approximation is illustrated in \cref{fig:weak_signal_approximation}. For the example parameters $\eta_0=0.8$, $\eta_{\pm1}=0.1$, and $\theta=2.5$, the approximation deviates by less than $1\%$ for $\lambda\lesssim0.2$. Within the range $\lambda\lesssim6$, the estimate remains conservative.

Using the typical values reported in \cref{tab:combined_eta_H} (Tukey window), the ratio of the average spectral leakage factor in the central bin to that in the first neighbor is $\hat\eta_0/\hat\eta_1\approx5$. Combining this with the estimate $|n/m| \approx 5\%$ at $\theta = 2.5$, neglecting the contribution from neighbouring bins, as done in previous works, yields an overestimated reslut of $\mu_{\rm CR}$ with a relative deviation of order $2\%$. This estimate is consistent with the results shown in \cref{fig:CR_distribution}.


\subsection{Threshold-only selection}
\label{sec:threshold_only}

If peaks are selected solely by exceeding the threshold $\theta$ (i.e.\ local-maxima selection is not applied), the peak-selection probability reduces to the simple tail probability of the non-central distribution,
\begin{equation}
    p_\lambda = \int_\theta^{+\infty} p(x;\lambda)\,dx.
\end{equation}
Under the weak-signal approximation, this expression linearizes to
\begin{equation}
    p_\lambda = e^{-\theta}\left(1+\frac{\theta}{2}\lambda\right) + \mathcal{O}(\lambda^2),
    \label{eq:p_lambda_threshold_only}
\end{equation}
so that $p_0=e^{-\theta}$ and one may identify the first-order statistic
\begin{equation}
    \Lambda^{(i)}_k = \frac{\theta}{2}\lambda^{(i)}_k.
\end{equation}
With this definition, the results for $\mu_n, \sigma_n^2$ and the CR moments given in \cref{eq:mu_n_sigma_n_app,eq:muCR_sigmaCR_app} remain valid (with the appropriate substitution for $\Lambda^{(i)}_k$).

\section{Spectral Leakage}
\label{sec:leakage}

\subsection{Spectral leakage factor}
\label{sec:spectral_leakage_factor}

When we focus on the frequency band of the signal, the normalized power spectrum $\eta_k$ in \cref{eq:eta_k} is the spectral leakage factor. To make this role explicit, we adopt a shifted-index convention that anchors frequency indices to the instantaneous signal frequency.

Let $t_c$ denote the central time of the segment and define the segment center frequency by $f_c\equiv f(t_c)$. With the DFT bin width $f_{\mathrm{bin}} = 1/T_{\mathrm{DFT}}$, the signal center expressed in bin units is
\begin{equation}
    k_c \equiv \frac{f_c}{f_{\mathrm{bin}}} = f_c T_{\mathrm{DFT}},
\end{equation}
which in general is not an integer. We identify the bin that contains the signal by the integer index $\lfloor k_{c} \rceil$, where $\lfloor \,\cdot\, \rceil$ denotes the rounding convention adopted here (the special treatment of half-integers is discussed in \cref{sec:Half-integer_kappa_case}).

Using $[k_c]$ as the reference, we introduce the signal-anchored index
\begin{equation}
    \kappa \equiv k - \lfloor k_{c} \rceil,
\end{equation}
so that $\kappa=0$ corresponds to the bin deemed to contain the signal and $\kappa=\pm 1$ label the immediate neighbors. In the shifted notation we write the leakage factors as $\eta_\kappa$, with $\eta_0$ giving the leakage into the signal bin and $\eta_{\pm1}$ the leakage into the neighbor bins. 

To characterize how spectral power is distributed across bins, we introduce a continuous spectral leakage function $\eta(o)$ defined on the offset coordinate
\begin{equation}
    o \equiv \frac{f - f_c}{f_{\mathrm{bin}}}\,,
    \label{eq:offset_factor_text}
\end{equation}
which measures frequency offsets in units of DFT bins. For a discrete frequency grid $f_k = k f_{\mathrm{bin}}$, one has the discrete offset
\begin{equation}
    o_k = \frac{f_k - f_c}{f_{\mathrm{bin}}} = k - k_c.
    \label{eq:offset_factor_grid}
\end{equation}
Using $k=\lfloor k_{c} \rceil+\kappa$, the offset relative to the anchored grid point is
\begin{equation}
    o_\kappa = \kappa + \big(\lfloor k_{c} \rceil-k_c\big),
    \label{eq:o_kappa_text}
\end{equation}
and in particular the offset of the central bin satisfies $o_0=\lfloor k_{c} \rceil-k_c\in(-\frac{1}{2},\frac{1}{2})$ under the rounding convention (the half-integer boundary is treated in \cref{sec:Half-integer_kappa_case}). Hence $o_\kappa\in(\kappa-\frac{1}{2},\kappa+\frac{1}{2})$.

The degree to which the signal frequency evolves across the segment is quantified by the dimensionless widening factor
\begin{equation}
    w \equiv \dot{f}\,T_{\mathrm{DFT}}^2 \;=\; \frac{\Delta f}{f_{\mathrm{bin}}},
    \label{eq:widening_factor}
\end{equation}
where $\Delta f=\dot f\,T_{\mathrm{DFT}}$ is the range of frequency evolution across the segment. If $w\ll 1$ the signal is approximately monochromatic within the segment; when $w\gtrsim \mathcal{O}(1)$ the frequency evolution across the STFT window must be accounted for explicitly.

In the monochromatic limit ($w\to 0$), the spectral leakage function reduces to the square of the modulus of the window's Fourier transform,
\begin{equation}
    \eta(o) = \left|\widetilde{W}(o)\right|^2,
    \label{eq:leakage_function}
\end{equation}
which is an even function of $o$. The discrete spectral leakage factors used in the main text are obtained by sampling this continuous function at the offsets $o_\kappa$:
\begin{equation}
    \eta_\kappa =\eta(o_\kappa) = \eta\big(o_0+\kappa\big),
    \label{eq:eta_sampling}
\end{equation}
where $o_0=\lfloor k_{c} \rceil-k_c$ fixes the sampling phase within the bin. Thus $\{\eta_\kappa\}$ are the equally spaced samples (spacing = 1 bin) of the underlying leakage profile; the sampling origin $o_0$ determines the relative power captured by the central and neighboring bins. The half-integer $\kappa$ cases are discussed in \cref{sec:Half-integer_kappa_case}.

\subsection{Spectral leakage function} 

\begin{table*}[hbt]
    \caption{Spectral leakage function under the monochromatic limit}
    \centering
    \renewcommand{\arraystretch}{2.5}
    \begin{tabular}{l l}
    \hline\hline
    Window             & Formula \\ 
    \hline
    Rectangle            & $\dfrac{\sin^2(\pi o)}{(\pi o)^2}$ \\
    Tukey ($\alpha = 0.5$)           & $\dfrac{256}{11} \dfrac{\sin^2(\frac{3\pi}{4} o)\cos^2(\frac{\pi}{4} o)}{(\pi o)^2(o^2-4)^2}$ \\
    Hanning            & $\dfrac{2}{3}\dfrac{1}{(o^2-1)^2}\dfrac{\sin^2(\pi o)}{(\pi o)^2}$ \\
    Hamming         & $\dfrac{0.733770 -0.217413 o^2+0.0161047 o^4}{(o^2-1)^2}\dfrac{\sin^2(\pi o)}{(\pi o)^2}$ \\
    Bartlett        & $\dfrac{12 \sin ^4(\pi o / 2)}{(\pi o)^4}$ \\
    Blackman        & $\dfrac{9.26592 -1.98555 o^2+0.106369  o^4}{(o^2-1)^2(o^2-4)^2}\dfrac{\sin^2(\pi o)}{(\pi o)^2}$ \\
    \hline\hline
    \end{tabular}
    \label{tab:window_formulas}
\end{table*}

To derive the form of the spectral leakage function 
$\eta(o)$ in the monochromatic limit, we consider a monochromatic signal of unit amplitude. In its continuous and discrete forms, the signal is given by
\begin{equation}
    \begin{aligned}
    s(t)&=e^{{i}(\phi_0+2\pi f t)},\\
    s[m]&=s\left(m\frac{T}{M}\right)=e^{i(\phi_0 + 2\pi m\frac{T}{M} f )}.
\end{aligned}
\end{equation}
The DFT power spectrum of this signal is
\begin{equation}
    \begin{aligned}
    \tilde{s}[k]&=\frac{e^{i\phi_0}}{M}\sum_{m=0}^{M-1} e^{i\frac{2\pi}{M}m fT  } e^{-i\frac{2 \pi}{M} mk}\\
    &=\frac{e^{i\phi_0}}{M}\sum_{m=0}^{M-1} \exp\left({i\frac{2\pi}{M} (fT -k)m}\right)\\
    &=\frac{e^{i\phi_0}}{M}\sum_{m=0}^{M-1} \exp\left({-i\frac{2\pi}{M} o_km}\right)\\
\end{aligned}
\label{eq:dft_mono}
\end{equation}
where we have used the discrete offset factor $o_k = k - f T$ from \cref{eq:offset_factor_grid}. Summing the geometric series in \cref{eq:dft_mono} yields
\begin{equation}
\tilde{s}[k] =
\begin{cases}
e^{i\phi_0} ~ \delta_{o_k,0}\,, & o_k=\text{integer},\\[12pt]
\frac{e^{i\phi_0}}{M}\frac{1 - \exp({-i 2\pi o_k})}{1 - \exp({-i 2\pi o_k/M})}\,, & o_k\neq\text{integer}.
\end{cases}
\label{eq:dft_mono_summed}
\end{equation}
When $f$ is an integer multiple of $1/T$, precisely aligned with the frequency grid, the spectrum  $|\tilde{s}[k]|^2=\delta_{fT,k}$ is a $\delta$ function at the corresponding frequency grid. If $f$ is not aligned with the frequency grid, then in the continuum limit ($M \to \infty$ with $T_{\mathrm{DFT}}$ fixed), 
\begin{equation}
    \lim_{M\to \infty}\tilde{s}[k]=\lim_{M\to \infty}\frac{e^{i\phi_0}}{M}\frac{1 - e^{-i2\pi o_k}}{1 - e^{-i2\pi o_k/M}}=e^{i\phi_0}\frac{1-e^{-i2\pi o_k}}{i2\pi o_k}
\end{equation}
The normalized power spectrum, which defines the spectral leakage function $\eta(o)$, is the squared magnitude of this result. For a rectangular window (implicitly used above), this gives
\begin{equation}
    \eta_\text{rect}(o)=\frac{\sin^2(\pi o)}{(\pi o)^2}.
\end{equation}

This result generalizes to an arbitrary window function $W(t)$ applied during the STFT. The spectral-leakage function is the squared magnitude of the window's normalized Fourier transform
\begin{equation}
    \eta(o) = \left| \frac{1}{T} \int_{0}^{T} W(t) ~e^{-i2\pi\frac{t}{T}o}dt \right|^2 = \left| \widetilde{W}(o) \right|^2.
    \label{eq:leakage_function_monochromatic}
\end{equation}
where $\widetilde{W}(o)$ is the Fourier transform evaluated at the offset factor $o$. The window function must satisfy the power normalization condition
\begin{equation}
    1=\frac{1}{T} \int_{0}^{T} |W(t) |^2dt=\frac{1}{M}\sum_{m=0}^{M-1}|W[m]|^2.
\end{equation}
The leakage functions $\eta(o)$ for several common window functions are listed in \cref{tab:window_formulas}.

\subsection{The case of half-integer \texorpdfstring{$k_c$}{kc}}
\label{sec:Half-integer_kappa_case}

A subtle point arises in the monochromatic limit ($w \to 0$) when the signal's center frequency $f_c$ falls exactly halfway between two DFT bins. In this case, the bin-scaled frequency $k_c = f_c T_{\mathrm{DFT}}$ is a half-integer (e.g., $n + 1/2$ for some integer $n$). This creates an ambiguity, as the signal is equidistant from bins $n$ and $n+1$, and neither can be uniquely designated as the ``signal bin''.

To handle this degenerate case formally, we adopt a special rounding convention for $\lfloor \,\cdot\, \rceil$ in which half-integers are preserved. That is, if $k_c = n + 1/2$, then we define $\lfloor k_{c} \rceil \equiv k_c$. Consequently, the anchored index $\kappa = k - \lfloor k_{c} \rceil$ will also be a half-integer for any integer bin index $k$. For the two bins adjacent to the true signal frequency, the anchored indices are
\begin{equation}
\begin{aligned}
    &\kappa = n - (n+1/2) = -1/2,\\
    &\kappa = (n+1) - (n+1/2) = +1/2.
\end{aligned}
\end{equation}
We may think of these half-integer indices as indicating that the two bins share the status of being ``closest'' to the signal.

In the monochromatic limit, the spectral-leakage function $\eta(o)$ is an even function of the offset $o$. Because our special convention implies $o_\kappa = \kappa + (\lfloor k_{c} \rceil - k_c) = \kappa$, the leakage factors for the two central bins become
\begin{equation}
\begin{aligned}
    &\eta_{-1/2} = \eta(o_{-1/2}) = \eta(-0.5), \\
    &\eta_{+1/2} = \eta(o_{+1/2}) = \eta(0.5).
\end{aligned}
\end{equation}
By symmetry, these two factors are equal: $\eta_{-1/2} = \eta_{+1/2}$. This relationship extends to all corresponding bin pairs, i.e., $\eta_{-(n+1/2)} = \eta_{n+1/2}$ for any integer $n$.

Although this convention provides a consistent way to handle the half-integer case, it is important to note that this scenario represents a set of measure zero. In practice, the probability that a signal's frequency lies exactly halfway between two bins is negligible.

\subsection{Averaged spectral leakage factor}
\label{sec:averaged_slf}

As observed in \cref{fig:three_eta}, the spectral leakage factors $\eta_\kappa$ oscillate rapidly over adjacent time segments. This behaviour arises because the sub-bin offset, $o_0 = [k_c] - k_c$, varies from one segment to the next. When the signal's widening factor $w$ is not a simple rational number, the sequence of offsets $\{o_0^{(i)}\}$ can be assumed to sample the interval $[-1/2, 1/2]$ uniformly. Under the weak-signal approximation, the detection statistic is linear in the non-centrality parameter $\lambda$. This linearity is crucial, as it allows the average of the statistic over the entire track to be expressed in terms of the average of $\lambda$, thereby justifying the replacement of the rapidly oscillating leakage factors with their mean values.

We define the averaged spectral leakage function $\hat{\eta}(o)$ as the moving average of $\eta(o)$ over one bin width:
\begin{equation}
    \hat{\eta}(o) \equiv \int_{o-1/2}^{o+1/2} \eta(o') \, \mathrm{d}o'.
\end{equation}
The averaged spectral leakage factors $\hat{\eta}_\kappa$, are then obtained by sampling this continuous function at integer offsets:
\begin{equation}
    \hat{\eta}_\kappa \equiv \hat{\eta}(o=\kappa).
    \label{eq:Average_SLF_revised}
\end{equation}
Since $\eta(o)$ is an even function of $o$, $\hat{\eta}(o)$ is also even, so we need only consider $\kappa \ge 0$. Values for common window functions are provided in \cref{tab:combined_eta_H}. \cref{fig:three_eta_move_average} confirms that applying a moving average to the data from \cref{fig:three_eta} converges to the values predicted by \cref{eq:Average_SLF_revised}.

This averaging procedure simplifies the calculation of the mean of the weak-signal statistic, $\bar{\Lambda}$, which appears in the distributions of the peak count $n$ and the CR (\cref{sec:weak_signal_approximation}). The track-averaged $\bar{\Lambda}$ is
\begin{equation}
\begin{aligned}
    \bar{\Lambda} &= \frac{1}{N} \sum_{i=1}^{N} \left[ m\lambda_k^{(i)} + n(\lambda_{k+1}^{(i)} + \lambda_{k-1}^{(i)}) \right] \\
    &\approx \frac{1}{N} \sum_{i=1}^{N} \mathcal{L}_i \left[ m\eta_k^{(i)} + n(\eta_{k+1}^{(i)} + \eta_{k-1}^{(i)}) \right].
\end{aligned}
\end{equation}
Here, we have factored out the total-power term $\mathcal{L}_i$, assuming that the noise PSD $S_n(f)$ is nearly constant over the few bins affected by leakage. If we further assume that $\mathcal{L}_i$ varies slowly compared to the oscillations in $\eta_\kappa^{(i)}$, we can approximate the average of the product as the product of the averages:
\begin{equation}
    \bar{\Lambda} \approx \frac{1}{N} \sum_{i=1}^{N} \mathcal{L}_i \left[ m\hat{\eta}_k + n(\hat{\eta}_{k+1} + \hat{\eta}_{k-1}) \right].
    \label{eq:bar_Lambda_eta_derivation_revised}
\end{equation}
This approximation effectively replaces the instantaneous leakage factors with their averaged counterparts in each segment.

To formalize this, we define an averaged combined spectral-leakage function, $\hat{\mathrm{H}}(o)$, and its corresponding factors, $\hat{\mathrm{H}}_\kappa$:
\begin{equation}
\begin{aligned}
    &\hat{\mathrm{H}}(o) \equiv m\hat{\eta}(o) + n[\hat{\eta}(o+1) + \hat{\eta}(o-1)], \\
    &\hat{\mathrm{H}}_\kappa \equiv \hat{\mathrm{H}}(o=\kappa) = m\hat{\eta}_\kappa + n(\hat{\eta}_{\kappa+1} + \hat{\eta}_{\kappa-1}).
    \label{eq:Eta_hat_revised}
\end{aligned}
\end{equation}
The track-averaged statistic $\bar{\Lambda}$ can now be expressed more generally. For a search track whose central frequency in the $i$-th segment is $f_{c,\mathrm{target}}^{(i)}$, offset from the real signal's central frequency $f_{c,\mathrm{real}}^{(i)}$ by
\begin{equation}
    o^{(i)} = \frac{f_{c,\mathrm{target}}^{(i)} - f_{c,\mathrm{real}}^{(i)}}{f_{\mathrm{bin}}},
\end{equation}
the averaged statistic is
\begin{equation}
    \bar{\Lambda} \approx \frac{1}{N} \sum_{i=1}^{N} \mathcal{L}_i \hat{\mathrm{H}}(o^{(i)}).
    \label{eq:bar_Lambda_Eta_general}
\end{equation}

This general form has a important special case. When searching on the real signal track, the target and real frequencies coincide, so $o^{(i)}=0$ for all $i$. \cref{eq:bar_Lambda_Eta_general} then simplifies to
\begin{equation}
    \bar{\Lambda}_{\text{on-track}} \approx \hat{\mathrm{H}}_0 \left( \frac{1}{N} \sum_{i=1}^{N} \mathcal{L}_i \right).
    \label{eq:bar_Lambda_on_track}
\end{equation}
A fixed integer offset, such as $\kappa=1$ in all segments, is generally unphysical because the corresponding time-frequency track does not match any real signal parameters, except for trivial cases like linearly evolving signals. Therefore, \cref{eq:bar_Lambda_Eta_general} is the appropriate form for modelling the response to mismatched search templates.

Finally, it is important to clarify the context of these spectral-leakage factors. The derivation, based on a complex exponential signal, corresponds to a single-sided spectrum where power is concentrated at a single frequency. For a real-valued signal, the signal power is symmetrically distributed between positive and negative frequencies. Therefore, when applying these results in the context of a double-sided spectrum for a real signal, the leakage factors defined here must be multiplied by a factor of $1/2$ to account for this distribution of power.

\bibliographystyle{utphys}
\bibliography{miniEMRI}

\providecommand{\href}[2]{#2}\begingroup\raggedright\begin{thebibliography}{10}

\bibitem{LIGOScientific:2018mvr}
{\bfseries LIGO Scientific, Virgo} Collaboration, B.~P. Abbott {\em et~al.}, ``{GWTC-1: A Gravitational-Wave Transient Catalog of Compact Binary Mergers Observed by LIGO and Virgo during the First and Second Observing Runs},'' \href{http://dx.doi.org/10.1103/PhysRevX.9.031040}{{\em Phys. Rev. X} {\bfseries 9} no.~3, (2019) 031040}, \href{http://arxiv.org/abs/1811.12907}{{\ttfamily arXiv:1811.12907 [astro-ph.HE]}}.

\bibitem{LIGOScientific:2020ibl}
{\bfseries LIGO Scientific, Virgo} Collaboration, R.~Abbott {\em et~al.}, ``{GWTC-2: Compact Binary Coalescences Observed by LIGO and Virgo During the First Half of the Third Observing Run},'' \href{http://dx.doi.org/10.1103/PhysRevX.11.021053}{{\em Phys. Rev. X} {\bfseries 11} (2021) 021053}, \href{http://arxiv.org/abs/2010.14527}{{\ttfamily arXiv:2010.14527 [gr-qc]}}.

\bibitem{KAGRA:2021vkt}
{\bfseries KAGRA, VIRGO, LIGO Scientific} Collaboration, R.~Abbott {\em et~al.}, ``{GWTC-3: Compact Binary Coalescences Observed by LIGO and Virgo during the Second Part of the Third Observing Run},'' \href{http://dx.doi.org/10.1103/PhysRevX.13.041039}{{\em Phys. Rev. X} {\bfseries 13} no.~4, (2023) 041039}, \href{http://arxiv.org/abs/2111.03606}{{\ttfamily arXiv:2111.03606 [gr-qc]}}.

\bibitem{LIGOScientific:2016aoc}
{\bfseries LIGO Scientific, Virgo} Collaboration, B.~P. Abbott {\em et~al.}, ``{Observation of Gravitational Waves from a Binary Black Hole Merger},'' \href{http://dx.doi.org/10.1103/PhysRevLett.116.061102}{{\em Phys. Rev. Lett.} {\bfseries 116} no.~6, (2016) 061102}, \href{http://arxiv.org/abs/1602.03837}{{\ttfamily arXiv:1602.03837 [gr-qc]}}.

\bibitem{LIGOScientific:2017vwq}
{\bfseries LIGO Scientific, Virgo} Collaboration, B.~P. Abbott {\em et~al.}, ``{GW170817: Observation of Gravitational Waves from a Binary Neutron Star Inspiral},'' \href{http://dx.doi.org/10.1103/PhysRevLett.119.161101}{{\em Phys. Rev. Lett.} {\bfseries 119} no.~16, (2017) 161101}, \href{http://arxiv.org/abs/1710.05832}{{\ttfamily arXiv:1710.05832 [gr-qc]}}.

\bibitem{LIGOScientific:2020aai}
{\bfseries LIGO Scientific, Virgo} Collaboration, B.~P. Abbott {\em et~al.}, ``{GW190425: Observation of a Compact Binary Coalescence with Total Mass $\sim 3.4 M_{\odot}$},'' \href{http://dx.doi.org/10.3847/2041-8213/ab75f5}{{\em Astrophys. J. Lett.} {\bfseries 892} no.~1, (2020) L3}, \href{http://arxiv.org/abs/2001.01761}{{\ttfamily arXiv:2001.01761 [astro-ph.HE]}}.

\bibitem{LIGOScientific:2020zkf}
{\bfseries LIGO Scientific, Virgo} Collaboration, R.~Abbott {\em et~al.}, ``{GW190814: Gravitational Waves from the Coalescence of a 23 Solar Mass Black Hole with a 2.6 Solar Mass Compact Object},'' \href{http://dx.doi.org/10.3847/2041-8213/ab960f}{{\em Astrophys. J. Lett.} {\bfseries 896} no.~2, (2020) L44}, \href{http://arxiv.org/abs/2006.12611}{{\ttfamily arXiv:2006.12611 [astro-ph.HE]}}.

\bibitem{LIGOScientific:2021qlt}
{\bfseries LIGO Scientific, KAGRA, VIRGO} Collaboration, R.~Abbott {\em et~al.}, ``{Observation of Gravitational Waves from Two Neutron Star\textendash{}Black Hole Coalescences},'' \href{http://dx.doi.org/10.3847/2041-8213/ac082e}{{\em Astrophys. J. Lett.} {\bfseries 915} no.~1, (2021) L5}, \href{http://arxiv.org/abs/2106.15163}{{\ttfamily arXiv:2106.15163 [astro-ph.HE]}}.

\bibitem{LIGOScientific:2020gml}
{\bfseries LIGO Scientific, Virgo} Collaboration, R.~Abbott {\em et~al.}, ``{Gravitational-wave Constraints on the Equatorial Ellipticity of Millisecond Pulsars},'' \href{http://dx.doi.org/10.3847/2041-8213/abb655}{{\em Astrophys. J. Lett.} {\bfseries 902} no.~1, (2020) L21}, \href{http://arxiv.org/abs/2007.14251}{{\ttfamily arXiv:2007.14251 [astro-ph.HE]}}.

\bibitem{LIGOScientific:2021hvc}
{\bfseries LIGO Scientific, VIRGO, KAGRA} Collaboration, R.~Abbott {\em et~al.}, ``{Searches for Gravitational Waves from Known Pulsars at Two Harmonics in the Second and Third LIGO-Virgo Observing Runs},'' \href{http://dx.doi.org/10.3847/1538-4357/ac6acf}{{\em Astrophys. J.} {\bfseries 935} no.~1, (2022) 1}, \href{http://arxiv.org/abs/2111.13106}{{\ttfamily arXiv:2111.13106 [astro-ph.HE]}}.

\bibitem{LIGOScientific:2021quq}
{\bfseries LIGO Scientific, KAGRA, VIRGO} Collaboration, R.~Abbott {\em et~al.}, ``{Narrowband Searches for Continuous and Long-duration Transient Gravitational Waves from Known Pulsars in the LIGO-Virgo Third Observing Run},'' \href{http://dx.doi.org/10.3847/1538-4357/ac6ad0}{{\em Astrophys. J.} {\bfseries 932} no.~2, (2022) 133}, \href{http://arxiv.org/abs/2112.10990}{{\ttfamily arXiv:2112.10990 [gr-qc]}}.

\bibitem{KAGRA:2022dwb}
{\bfseries KAGRA, LIGO Scientific, VIRGO} Collaboration, R.~Abbott {\em et~al.}, ``{All-sky search for continuous gravitational waves from isolated neutron stars using Advanced LIGO and Advanced Virgo O3 data},'' \href{http://dx.doi.org/10.1103/PhysRevD.106.102008}{{\em Phys. Rev. D} {\bfseries 106} no.~10, (2022) 102008}, \href{http://arxiv.org/abs/2201.00697}{{\ttfamily arXiv:2201.00697 [gr-qc]}}.

\bibitem{KAGRA:2022osp}
{\bfseries KAGRA, LIGO Scientific, VIRGO} Collaboration, R.~Abbott {\em et~al.}, ``{Search for continuous gravitational wave emission from the Milky~Way center in O3 LIGO-Virgo data},'' \href{http://dx.doi.org/10.1103/PhysRevD.106.042003}{{\em Phys. Rev. D} {\bfseries 106} no.~4, (2022) 042003}, \href{http://arxiv.org/abs/2204.04523}{{\ttfamily arXiv:2204.04523 [astro-ph.HE]}}.

\bibitem{Arvanitaki:2014wva}
A.~Arvanitaki, M.~Baryakhtar, and X.~Huang, ``{Discovering the QCD Axion with Black Holes and Gravitational Waves},'' \href{http://dx.doi.org/10.1103/PhysRevD.91.084011}{{\em Phys. Rev. D} {\bfseries 91} no.~8, (2015) 084011}, \href{http://arxiv.org/abs/1411.2263}{{\ttfamily arXiv:1411.2263 [hep-ph]}}.

\bibitem{Brito:2015oca}
R.~Brito, V.~Cardoso, and P.~Pani, ``{Superradiance}: {New Frontiers in Black Hole Physics},'' \href{http://dx.doi.org/10.1007/978-3-319-19000-6}{{\em Lect. Notes Phys.} {\bfseries 906} (2015) pp.1--237}, \href{http://arxiv.org/abs/1501.06570}{{\ttfamily arXiv:1501.06570 [gr-qc]}}.

\bibitem{LIGOScientific:2021rnv}
{\bfseries LIGO Scientific, Virgo, KAGRA} Collaboration, R.~Abbott {\em et~al.}, ``{All-sky search for gravitational wave emission from scalar boson clouds around spinning black holes in LIGO O3 data},'' \href{http://dx.doi.org/10.1103/PhysRevD.105.102001}{{\em Phys. Rev. D} {\bfseries 105} no.~10, (2022) 102001}, \href{http://arxiv.org/abs/2111.15507}{{\ttfamily arXiv:2111.15507 [astro-ph.HE]}}.

\bibitem{Pierce:2018xmy}
A.~Pierce, K.~Riles, and Y.~Zhao, ``{Searching for Dark Photon Dark Matter with Gravitational Wave Detectors},'' \href{http://dx.doi.org/10.1103/PhysRevLett.121.061102}{{\em Phys. Rev. Lett.} {\bfseries 121} no.~6, (2018) 061102}, \href{http://arxiv.org/abs/1801.10161}{{\ttfamily arXiv:1801.10161 [hep-ph]}}.

\bibitem{Guo:2019ker}
H.-K. Guo, K.~Riles, F.-W. Yang, and Y.~Zhao, ``{Searching for Dark Photon Dark Matter in LIGO O1 Data},'' \href{http://dx.doi.org/10.1038/s42005-019-0255-0}{{\em Commun. Phys.} {\bfseries 2} (2019) 155}, \href{http://arxiv.org/abs/1905.04316}{{\ttfamily arXiv:1905.04316 [hep-ph]}}.

\bibitem{Miller:2020vsl}
A.~L. Miller {\em et~al.}, ``{Probing new light gauge bosons with gravitational-wave interferometers using an adapted semicoherent method},'' \href{http://dx.doi.org/10.1103/PhysRevD.103.103002}{{\em Phys. Rev. D} {\bfseries 103} no.~10, (2021) 103002}, \href{http://arxiv.org/abs/2010.01925}{{\ttfamily arXiv:2010.01925 [astro-ph.IM]}}.

\bibitem{Vermeulen:2021epa}
S.~M. Vermeulen {\em et~al.}, ``{Direct limits for scalar field dark matter from a gravitational-wave detector},'' \href{http://arxiv.org/abs/2103.03783}{{\ttfamily arXiv:2103.03783 [gr-qc]}}.

\bibitem{LIGOScientific:2021ffg}
{\bfseries LIGO Scientific, KAGRA, Virgo} Collaboration, R.~Abbott {\em et~al.}, ``{Constraints on dark photon dark matter using data from LIGO\textquoteright{}s and Virgo\textquoteright{}s third observing run},'' \href{http://dx.doi.org/10.1103/PhysRevD.105.063030}{{\em Phys. Rev. D} {\bfseries 105} no.~6, (2022) 063030}, \href{http://arxiv.org/abs/2105.13085}{{\ttfamily arXiv:2105.13085 [astro-ph.CO]}}. [Erratum: Phys.Rev.D 109, 089902 (2024)].

\bibitem{Miller:2022wxu}
{\bfseries LIGO Scientific Collaboration, Virgo, KAGRA} Collaboration, A.~L. Miller, F.~Badaracco, and C.~Palomba, ``{Distinguishing between dark-matter interactions with gravitational-wave detectors},'' \href{http://dx.doi.org/10.1103/PhysRevD.105.103035}{{\em Phys. Rev. D} {\bfseries 105} no.~10, (2022) 103035}, \href{http://arxiv.org/abs/2204.03814}{{\ttfamily arXiv:2204.03814 [astro-ph.IM]}}.

\bibitem{Manita:2023mnc}
Y.~Manita, H.~Takeda, K.~Aoki, T.~Fujita, and S.~Mukohyama, ``{Exploring the spin of ultralight dark matter with gravitational wave detectors},'' \href{http://dx.doi.org/10.1103/PhysRevD.109.095012}{{\em Phys. Rev. D} {\bfseries 109} no.~9, (2024) 095012}, \href{http://arxiv.org/abs/2310.10646}{{\ttfamily arXiv:2310.10646 [hep-ph]}}.

\bibitem{KAGRA:2024ipf}
{\bfseries KAGRA, LIGO Scientific, VIRGO} Collaboration, A.~G. Abac {\em et~al.}, ``{Ultralight vector dark matter search using data from the KAGRA O3GK run},'' \href{http://dx.doi.org/10.1103/PhysRevD.110.042001}{{\em Phys. Rev. D} {\bfseries 110} no.~4, (2024) 042001}, \href{http://arxiv.org/abs/2403.03004}{{\ttfamily arXiv:2403.03004 [astro-ph.CO]}}.

\bibitem{LIGOScientific:2025ttj}
{\bfseries LIGO Scientific, VIRGO, KAGRA} Collaboration, A.~G. Abac {\em et~al.}, ``{Direct multi-model dark-matter search with gravitational-wave interferometers using data from the first part of the fourth LIGO-Virgo-KAGRA observing run},'' \href{http://arxiv.org/abs/2510.27022}{{\ttfamily arXiv:2510.27022 [astro-ph.CO]}}.

\bibitem{LIGOScientific:2018glc}
{\bfseries LIGO Scientific Collaboration, Virgo} Collaboration, B.~P. Abbott {\em et~al.}, ``{Search for Subsolar-Mass Ultracompact Binaries in Advanced LIGO\textquoteright{}s First Observing Run},'' \href{http://dx.doi.org/10.1103/PhysRevLett.121.231103}{{\em Phys. Rev. Lett.} {\bfseries 121} no.~23, (2018) 231103}, \href{http://arxiv.org/abs/1808.04771}{{\ttfamily arXiv:1808.04771 [astro-ph.CO]}}.

\bibitem{LIGOScientific:2019kan}
{\bfseries LIGO Scientific, Virgo} Collaboration, B.~P. Abbott {\em et~al.}, ``{Search for Subsolar Mass Ultracompact Binaries in Advanced LIGO\textquoteright{}s Second Observing Run},'' \href{http://dx.doi.org/10.1103/PhysRevLett.123.161102}{{\em Phys. Rev. Lett.} {\bfseries 123} no.~16, (2019) 161102}, \href{http://arxiv.org/abs/1904.08976}{{\ttfamily arXiv:1904.08976 [astro-ph.CO]}}.

\bibitem{Miller:2020kmv}
A.~L. Miller, S.~Clesse, F.~De~Lillo, G.~Bruno, A.~Depasse, and A.~Tanasijczuk, ``{Probing planetary-mass primordial black holes with continuous gravitational waves},'' \href{http://dx.doi.org/10.1016/j.dark.2021.100836}{{\em Phys. Dark Univ.} {\bfseries 32} (2021) 100836}, \href{http://arxiv.org/abs/2012.12983}{{\ttfamily arXiv:2012.12983 [astro-ph.HE]}}.

\bibitem{LIGOScientific:2021job}
{\bfseries LIGO Scientific Collaboration, Virgo, KAGRA} Collaboration, R.~Abbott {\em et~al.}, ``{Search for Subsolar-Mass Binaries in the First Half of Advanced LIGO\textquoteright{}s and Advanced Virgo\textquoteright{}s Third Observing Run},'' \href{http://dx.doi.org/10.1103/PhysRevLett.129.061104}{{\em Phys. Rev. Lett.} {\bfseries 129} no.~6, (2022) 061104}, \href{http://arxiv.org/abs/2109.12197}{{\ttfamily arXiv:2109.12197 [astro-ph.CO]}}.

\bibitem{Nitz:2021mzz}
A.~H. Nitz and Y.-F. Wang, ``Search for gravitational waves from the coalescence of subsolar mass and eccentric compact binaries,'' \href{http://dx.doi.org/10.3847/1538-4357/ac01d9}{{\em The Astrophysical Journal} {\bfseries 915} no.~1, (Jul, 2021) 54}, \href{http://arxiv.org/abs/arXiv:2102.00868}{{\ttfamily arXiv:2102.00868}}. \url{https://dx.doi.org/10.3847/1538-4357/ac01d9}.

\bibitem{LIGOScientific:2022hai}
{\bfseries LIGO Scientific, VIRGO, KAGRA} Collaboration, R.~Abbott {\em et~al.}, ``{Search for subsolar-mass black hole binaries in the second part of Advanced LIGO's and Advanced Virgo's third observing run},'' \href{http://dx.doi.org/10.1093/mnras/stad588}{{\em Mon. Not. Roy. Astron. Soc.} {\bfseries 524} no.~4, (2023) 5984--5992}, \href{http://arxiv.org/abs/2212.01477}{{\ttfamily arXiv:2212.01477 [astro-ph.HE]}}. [Erratum: Mon.Not.Roy.Astron.Soc. 526, 6234 (2023)].

\bibitem{Nitz:2022ltl}
A.~H. Nitz and Y.-F. Wang, ``{Broad search for gravitational waves from subsolar-mass binaries through LIGO and Virgo\textquoteright{}s third observing run},'' \href{http://dx.doi.org/10.1103/PhysRevD.106.023024}{{\em Phys. Rev. D} {\bfseries 106} no.~2, (2022) 023024}, \href{http://arxiv.org/abs/2202.11024}{{\ttfamily arXiv:2202.11024 [astro-ph.HE]}}.

\bibitem{Miller:2024rca}
A.~L. Miller, {\em Gravitational Waves from Sub-Solar Mass Primordial Black Holes}, \href{http://dx.doi.org/10.1007/978-981-97-8887-3_18}{pp.~467--494}.
\newblock Springer Nature Singapore, Singapore, 2025.
\newblock \url{https://doi.org/10.1007/978-981-97-8887-3_18}.

\bibitem{Miller:2024fpo}
A.~L. Miller, N.~Aggarwal, S.~Clesse, F.~De~Lillo, S.~Sachdev, P.~Astone, C.~Palomba, O.~J. Piccinni, and L.~Pierini, ``{Gravitational Wave Constraints on Planetary-Mass Primordial Black Holes Using LIGO O3a Data},'' \href{http://dx.doi.org/10.1103/PhysRevLett.133.111401}{{\em Phys. Rev. Lett.} {\bfseries 133} no.~11, (2024) 111401}, \href{http://arxiv.org/abs/2402.19468}{{\ttfamily arXiv:2402.19468 [gr-qc]}}.

\bibitem{LIGOScientific:2025vwc}
{\bfseries LIGO Scientific Collaboration, VIRGO, KAGRA} Collaboration, A.~G. Abac {\em et~al.}, ``{Search for planetary-mass ultra-compact binaries using data from the first part of the LIGO--Virgo--KAGRA fourth observing run},'' \href{http://arxiv.org/abs/2511.19911}{{\ttfamily arXiv:2511.19911 [gr-qc]}}.

\bibitem{Piccinni:2022vsd}
O.~J. Piccinni, ``{Status and Perspectives of Continuous Gravitational Wave Searches},'' \href{http://dx.doi.org/10.3390/galaxies10030072}{{\em Galaxies} {\bfseries 10} no.~3, (2022) 72}, \href{http://arxiv.org/abs/2202.01088}{{\ttfamily arXiv:2202.01088 [gr-qc]}}.

\bibitem{Haskell:2023yrv}
B.~Haskell and M.~Bejger, ``{Astrophysics with continuous gravitational waves},'' \href{http://dx.doi.org/10.1038/s41550-023-02059-w}{{\em Nature Astron.} {\bfseries 7} no.~10, (2023) 1160--1170}.

\bibitem{Bertone:2019irm}
G.~Bertone {\em et~al.}, ``{Gravitational wave probes of dark matter: challenges and opportunities},'' \href{http://dx.doi.org/10.21468/SciPostPhysCore.3.2.007}{{\em SciPost Phys. Core} {\bfseries 3} (2020) 007}, \href{http://arxiv.org/abs/1907.10610}{{\ttfamily arXiv:1907.10610 [astro-ph.CO]}}.

\bibitem{Bertone:2024rxe}
G.~Bertone, ``{Dark matter, black holes, and gravitational waves},'' \href{http://dx.doi.org/10.1016/j.nuclphysb.2024.116487}{{\em Nucl. Phys. B} {\bfseries 1003} (2024) 116487}, \href{http://arxiv.org/abs/2404.11513}{{\ttfamily arXiv:2404.11513 [astro-ph.CO]}}.

\bibitem{Miller:2025yyx}
A.~L. Miller, ``{Gravitational wave probes of particle dark matter: a review},'' \href{http://arxiv.org/abs/2503.02607}{{\ttfamily arXiv:2503.02607 [astro-ph.HE]}}.

\bibitem{Zeldovich:1967lct}
Y.~B. Zel'dovich and I.~D. Novikov, ``{The Hypothesis of Cores Retarded during Expansion and the Hot Cosmological Model},'' {\em Sov. Astron.} {\bfseries 10} (1967) 602.

\bibitem{Hawking:1971ei}
S.~Hawking, ``{Gravitationally collapsed objects of very low mass},'' \href{http://dx.doi.org/10.1093/mnras/152.1.75}{{\em Mon. Not. Roy. Astron. Soc.} {\bfseries 152} (1971) 75}.

\bibitem{Carr:1974nx}
B.~J. Carr and S.~W. Hawking, ``{Black holes in the early Universe},'' \href{http://dx.doi.org/10.1093/mnras/168.2.399}{{\em Mon. Not. Roy. Astron. Soc.} {\bfseries 168} (1974) 399--415}.

\bibitem{Carr:2020gox}
B.~Carr, K.~Kohri, Y.~Sendouda, and J.~Yokoyama, ``{Constraints on primordial black holes},'' \href{http://dx.doi.org/10.1088/1361-6633/ac1e31}{{\em Rept. Prog. Phys.} {\bfseries 84} no.~11, (2021) 116902}, \href{http://arxiv.org/abs/2002.12778}{{\ttfamily arXiv:2002.12778 [astro-ph.CO]}}.

\bibitem{Carr:2020xqk}
B.~Carr and F.~Kuhnel, ``{Primordial Black Holes as Dark Matter: Recent Developments},'' \href{http://dx.doi.org/10.1146/annurev-nucl-050520-125911}{{\em Ann. Rev. Nucl. Part. Sci.} {\bfseries 70} (2020) 355--394}, \href{http://arxiv.org/abs/2006.02838}{{\ttfamily arXiv:2006.02838 [astro-ph.CO]}}.

\bibitem{Bird:2022wvk}
S.~Bird {\em et~al.}, ``{Snowmass2021 Cosmic Frontier White Paper: Primordial black hole dark matter},'' \href{http://dx.doi.org/10.1016/j.dark.2023.101231}{{\em Phys. Dark Univ.} {\bfseries 41} (2023) 101231}, \href{http://arxiv.org/abs/2203.08967}{{\ttfamily arXiv:2203.08967 [hep-ph]}}.

\bibitem{Niikura:2019kqi}
H.~Niikura, M.~Takada, S.~Yokoyama, T.~Sumi, and S.~Masaki, ``{Constraints on Earth-mass primordial black holes from OGLE 5-year microlensing events},'' \href{http://dx.doi.org/10.1103/PhysRevD.99.083503}{{\em Phys. Rev. D} {\bfseries 99} no.~8, (2019) 083503}, \href{http://arxiv.org/abs/1901.07120}{{\ttfamily arXiv:1901.07120 [astro-ph.CO]}}.

\bibitem{Bhatiani:2019}
S.~{Bhatiani}, X.~{Dai}, and E.~{Guerras}, ``{Confirmation of Planet-mass Objects in Extragalactic Systems},'' \href{http://dx.doi.org/10.3847/1538-4357/ab46ac}{{\em Astrophys. J.} {\bfseries 885} no.~1, (Nov., 2019) 77}, \href{http://arxiv.org/abs/1909.11610}{{\ttfamily arXiv:1909.11610 [astro-ph.GA]}}.

\bibitem{Hawkins:2020zie}
M.~R.~S. Hawkins, ``{The signature of primordial black holes in the dark matter halos of galaxies},'' \href{http://dx.doi.org/10.1051/0004-6361/201936462}{{\em Astron. Astrophys.} {\bfseries 633} (2020) A107}, \href{http://arxiv.org/abs/2001.07633}{{\ttfamily arXiv:2001.07633 [astro-ph.GA]}}.

\bibitem{Barsanti:2021ydd}
S.~Barsanti, V.~De~Luca, A.~Maselli, and P.~Pani, ``{Detecting Subsolar-Mass Primordial Black Holes in Extreme Mass-Ratio Inspirals with LISA and Einstein Telescope},'' \href{http://dx.doi.org/10.1103/PhysRevLett.128.111104}{{\em Phys. Rev. Lett.} {\bfseries 128} no.~11, (2022) 111104}, \href{http://arxiv.org/abs/2109.02170}{{\ttfamily arXiv:2109.02170 [gr-qc]}}.

\bibitem{Miller:2024khl}
A.~L. Miller, ``{Prospects for detecting asteroid-mass primordial black holes in extreme-mass-ratio inspirals with continuous gravitational waves},'' \href{http://dx.doi.org/10.1103/j1h1-hl62}{{\em Phys. Rev. D} {\bfseries 112} no.~10, (2025) 103027}, \href{http://arxiv.org/abs/2410.01348}{{\ttfamily arXiv:2410.01348 [gr-qc]}}.

\bibitem{Guo:2022sdd}
H.-K. Guo and A.~Miller, ``{Searching for Mini Extreme Mass Ratio Inspirals with Gravitational-Wave Detectors},'' \href{http://arxiv.org/abs/2205.10359}{{\ttfamily arXiv:2205.10359 [astro-ph.IM]}}.

\bibitem{LVK:2022ydq}
{\bfseries LVK} Collaboration, R.~Abbott {\em et~al.}, ``{Search for subsolar-mass black hole binaries in the second part of Advanced LIGO\textquoteright{}s and Advanced Virgo\textquoteright{}s third observing run},'' \href{http://dx.doi.org/10.1093/mnras/stad588}{{\em Mon. Not. Roy. Astron. Soc.} {\bfseries 524} no.~4, (2023) 5984--5992}, \href{http://arxiv.org/abs/2212.01477}{{\ttfamily arXiv:2212.01477 [astro-ph.HE]}}. [Erratum: Mon.Not.Roy.Astron.Soc. 526, 6234 (2023)].

\bibitem{Phukon:2021cus}
K.~S. Phukon, G.~Baltus, S.~Caudill, S.~Clesse, A.~Depasse, M.~Fays, H.~Fong, S.~J. Kapadia, R.~Magee, and A.~J. Tanasijczuk, ``The hunt for sub-solar primordial black holes in low mass ratio binaries is open,'' \href{http://arxiv.org/abs/2105.11449}{{\ttfamily arXiv:2105.11449 [astro-ph.CO]}}.

\bibitem{Alestas:2024ubs}
G.~Alestas, G.~Morras, T.~S. Yamamoto, J.~Garcia-Bellido, S.~Kuroyanagi, and S.~Nesseris, ``{Applying the Viterbi algorithm to planetary-mass black hole searches},'' \href{http://dx.doi.org/10.1103/PhysRevD.109.123516}{{\em Phys. Rev. D} {\bfseries 109} no.~12, (2024) 123516}, \href{http://arxiv.org/abs/2401.02314}{{\ttfamily arXiv:2401.02314 [astro-ph.CO]}}.

\bibitem{Andres-Carcasona:2023zny}
M.~Andr\'es-Carcasona, O.~J. Piccinni, M.~Mart\'\i{}nez, and L.-M. Mir, ``{BSD-COBI: New search pipeline to target inspiraling light dark compact objects.},'' \href{http://dx.doi.org/10.22323/1.449.0067}{{\em PoS} {\bfseries EPS-HEP2023} (2024) 067}.

\bibitem{Andres-Carcasona:2024jvz}
M.~Andr\'es-Carcasona, O.~J. Piccinni, M.~Mart\'\i{}nez, and L.~M. Mir, ``{New approach to search for long transient gravitational waves from inspiraling compact binary systems},'' \href{http://arxiv.org/abs/2411.04498}{{\ttfamily arXiv:2411.04498 [gr-qc]}}.

\bibitem{Tenorio:2025gci}
R.~Tenorio and D.~Gerosa, ``{Scalable data-analysis framework for long-duration gravitational waves from compact binaries using short Fourier transforms},'' \href{http://dx.doi.org/10.1103/PhysRevD.111.104044}{{\em Phys. Rev. D} {\bfseries 111} no.~10, (2025) 104044}, \href{http://arxiv.org/abs/2502.11823}{{\ttfamily arXiv:2502.11823 [gr-qc]}}.

\bibitem{Tenorio:2024jgc}
R.~Tenorio, J.-R. M{\'e}rou, and A.~M. Sintes, ``{One-stop strategy to search for long-duration gravitational-wave signals},'' \href{http://dx.doi.org/10.1103/PhysRevD.111.104002}{{\em Phys. Rev. D} {\bfseries 111} no.~10, (2025) 104002}, \href{http://arxiv.org/abs/2411.18370}{{\ttfamily arXiv:2411.18370 [gr-qc]}}.

\bibitem{Merou:2025ark}
J.-R. M{\'e}rou, R.~Tenorio, and A.~M. Sintes, ``{GPU-Accelerated Searches for Long-Transient Gravitational Waves from Newborn Neutron Stars},''
\newblock 7, 2025.
\newblock \href{http://arxiv.org/abs/2507.07816}{{\ttfamily arXiv:2507.07816 [gr-qc]}}.

\bibitem{Miller:2021knj}
A.~L. Miller, N.~Aggarwal, S.~Clesse, and F.~De~Lillo, ``{Constraints on planetary and asteroid-mass primordial black holes from continuous gravitational-wave searches},'' \href{http://dx.doi.org/10.1103/PhysRevD.105.062008}{{\em Phys. Rev. D} {\bfseries 105} no.~6, (2022) 062008}, \href{http://arxiv.org/abs/2110.06188}{{\ttfamily arXiv:2110.06188 [gr-qc]}}.

\bibitem{Miller:2024jpo}
A.~L. Miller, N.~Aggarwal, S.~Clesse, F.~De~Lillo, S.~Sachdev, P.~Astone, C.~Palomba, O.~J. Piccinni, and L.~Pierini, ``{Method to search for inspiraling planetary-mass ultracompact binaries using the generalized frequency-Hough transform in LIGO O3a data},'' \href{http://dx.doi.org/10.1103/PhysRevD.110.082004}{{\em Phys. Rev. D} {\bfseries 110} no.~8, (2024) 082004}, \href{http://arxiv.org/abs/2407.17052}{{\ttfamily arXiv:2407.17052 [astro-ph.IM]}}.

\bibitem{Krishnan:2004sv}
B.~Krishnan, A.~M. Sintes, M.~A. Papa, B.~F. Schutz, S.~Frasca, and C.~Palomba, ``{The Hough transform search for continuous gravitational waves},'' \href{http://dx.doi.org/10.1103/PhysRevD.70.082001}{{\em Phys. Rev. D} {\bfseries 70} (2004) 082001}, \href{http://arxiv.org/abs/gr-qc/0407001}{{\ttfamily arXiv:gr-qc/0407001}}.

\bibitem{Palomba:2005fp}
C.~Palomba, P.~Astone, and S.~Frasca, ``{Adaptive Hough transform for the search of periodic sources},'' \href{http://dx.doi.org/10.1088/0264-9381/22/18/S39}{{\em Class. Quant. Grav.} {\bfseries 22} (2005) S1255--S1264}.

\bibitem{Astone:2014esa}
P.~Astone, A.~Colla, S.~D'Antonio, S.~Frasca, and C.~Palomba, ``{Method for all-sky searches of continuous gravitational wave signals using the frequency-Hough transform},'' \href{http://dx.doi.org/10.1103/PhysRevD.90.042002}{{\em Phys. Rev. D} {\bfseries 90} no.~4, (2014) 042002}, \href{http://arxiv.org/abs/1407.8333}{{\ttfamily arXiv:1407.8333 [astro-ph.IM]}}.

\bibitem{Jaranowski:1998qm}
P.~Jaranowski, A.~Krolak, and B.~F. Schutz, ``{Data analysis of gravitational - wave signals from spinning neutron stars. 1. The Signal and its detection},'' \href{http://dx.doi.org/10.1103/PhysRevD.58.063001}{{\em Phys. Rev. D} {\bfseries 58} (1998) 063001}, \href{http://arxiv.org/abs/gr-qc/9804014}{{\ttfamily arXiv:gr-qc/9804014}}.

\bibitem{Riles:2022wwz}
K.~Riles, ``{Searches for continuous-wave gravitational radiation},'' \href{http://dx.doi.org/10.1007/s41114-023-00044-3}{{\em Living Rev. Rel.} {\bfseries 26} no.~1, (2023) 3}, \href{http://arxiv.org/abs/2206.06447}{{\ttfamily arXiv:2206.06447 [astro-ph.HE]}}.

\bibitem{Maggiore:2007ulw}
M.~Maggiore, \href{http://dx.doi.org/10.1093/acprof:oso/9780198570745.001.0001}{{\em {Gravitational Waves. Vol. 1: Theory and Experiments}}}.
\newblock Oxford University Press, 2007.

\bibitem{Astone:2005fj}
P.~Astone, S.~Frasca, and C.~Palomba, ``{The short FFT database and the peak map for the hierarchical search of periodic sources},'' \href{http://dx.doi.org/10.1088/0264-9381/22/18/S34}{{\em Class. Quant. Grav.} {\bfseries 22} (2005) S1197--S1210}.

\bibitem{Miller:2018rbg}
A.~Miller {\em et~al.}, ``{Method to search for long duration gravitational wave transients from isolated neutron stars using the generalized frequency-Hough transform},'' \href{http://dx.doi.org/10.1103/PhysRevD.98.102004}{{\em Phys. Rev. D} {\bfseries 98} no.~10, (2018) 102004}, \href{http://arxiv.org/abs/1810.09784}{{\ttfamily arXiv:1810.09784 [astro-ph.IM]}}.

\bibitem{Menon:2025wce}
S.~S. Menon {\em et~al.}, ``{GFH-v2 Pipeline for Searches of Long-Transient Gravitational Waves from Newborn Magnetars},'' \href{http://arxiv.org/abs/2512.09878}{{\ttfamily arXiv:2512.09878 [astro-ph.IM]}}.

\bibitem{Miller:2025ote}
A.~L. Miller and L.~Pierini, ``{BinaryGFH-v2: Improved method to search for gravitational waves from sub-solar-mass, ultra-compact binaries using the Generalized Frequency-Hough Transform},'' \href{http://arxiv.org/abs/2512.10539}{{\ttfamily arXiv:2512.10539 [gr-qc]}}.

\bibitem{Miller:2019jtp}
A.~L. Miller {\em et~al.}, ``{How effective is machine learning to detect long transient gravitational waves from neutron stars in a real search?},'' \href{http://dx.doi.org/10.1103/PhysRevD.100.062005}{{\em Phys. Rev. D} {\bfseries 100} no.~6, (2019) 062005}, \href{http://arxiv.org/abs/1909.02262}{{\ttfamily arXiv:1909.02262 [astro-ph.IM]}}.

\bibitem{Palomba:2025}
C.~Palomba, ``On the sensitivity of peakmap-based methods for the search of continuous gravitational wave signals,'' Aug., 2025.
\newblock \url{https://tds.virgo-gw.eu/?r=25138}.

\bibitem{Allen:2002bp}
B.~Allen, M.~A. Papa, and B.~F. Schutz, ``{Optimal strategies for sinusoidal signal detection},'' \href{http://dx.doi.org/10.1103/PhysRevD.66.102003}{{\em Phys. Rev. D} {\bfseries 66} (2002) 102003}, \href{http://arxiv.org/abs/gr-qc/0206032}{{\ttfamily arXiv:gr-qc/0206032}}.

\bibitem{harris2020array}
C.~R. Harris, K.~J. Millman, S.~J. van~der Walt, R.~Gommers, P.~Virtanen, D.~Cournapeau, E.~Wieser, J.~Taylor, S.~Berg, N.~J. Smith, R.~Kern, M.~Picus, S.~Hoyer, M.~H. van Kerkwijk, M.~Brett, A.~Haldane, J.~F. del R{\'{i}}o, M.~Wiebe, P.~Peterson, P.~G{\'{e}}rard-Marchant, K.~Sheppard, T.~Reddy, W.~Weckesser, H.~Abbasi, C.~Gohlke, and T.~E. Oliphant, ``Array programming with {NumPy},'' \href{http://dx.doi.org/10.1038/s41586-020-2649-2}{{\em Nature} {\bfseries 585} no.~7825, (Sept., 2020) 357--362}. \url{https://doi.org/10.1038/s41586-020-2649-2}.

\bibitem{2020SciPy-NMeth}
P.~Virtanen, R.~Gommers, T.~E. Oliphant, M.~Haberland, T.~Reddy, D.~Cournapeau, E.~Burovski, P.~Peterson, W.~Weckesser, J.~Bright, S.~J. {van der Walt}, M.~Brett, J.~Wilson, K.~J. Millman, N.~Mayorov, A.~R.~J. Nelson, E.~Jones, R.~Kern, E.~Larson, C.~J. Carey, {\.I}.~Polat, Y.~Feng, E.~W. Moore, J.~{VanderPlas}, D.~Laxalde, J.~Perktold, R.~Cimrman, I.~Henriksen, E.~A. Quintero, C.~R. Harris, A.~M. Archibald, A.~H. Ribeiro, F.~Pedregosa, P.~{van Mulbregt}, and {SciPy 1.0 Contributors}, ``{{SciPy} 1.0: Fundamental Algorithms for Scientific Computing in Python},'' \href{http://dx.doi.org/10.1038/s41592-019-0686-2}{{\em Nature Methods} {\bfseries 17} (2020) 261--272}.

\bibitem{Hunter:2007}
J.~D. Hunter, ``Matplotlib: A 2d graphics environment,'' \href{http://dx.doi.org/10.1109/MCSE.2007.55}{{\em Computing in Science \& Engineering} {\bfseries 9} no.~3, (2007) 90--95}.

\end{thebibliography}\endgroup

\end{document}